\documentclass[12pt]{article}

\pdfoutput=1

\usepackage[top=80pt,bottom=85pt,left=60pt,right=60pt]{geometry}
\usepackage{amssymb}
\usepackage{amsmath}
\usepackage{graphicx,color,subfigure}
\usepackage{float}
\usepackage{cite}
\usepackage{enumerate}
\usepackage[debug,pageanchor=false]{hyperref}
\definecolor{link}{rgb}{.8,.15,.1}
\hypersetup{colorlinks=true,linkcolor=link,citecolor=link,urlcolor=link,linktocpage}

\newcommand{\R}{\mathbb{R}}

\renewcommand{\b}{b}

\DeclareMathOperator{\e}{e}

\DeclareMathOperator{\eps}{\epsilon}

\makeatletter
\@addtoreset{equation}{section}
\makeatother

\newcommand{\beq}{\begin{equation}}
\newcommand{\eeq}{\end{equation}}
\newcommand{\bea}{\begin{eqnarray}}
\newcommand{\eea}{\end{eqnarray}}
\newcommand{\nn}{\nonumber}

\begin{document}

\begin{titlepage}

\begin{center}

\vskip .5in 
\noindent

{\Large \bf{Minkowski$_4\times S^2$ solutions of IIB supergravity}}

\bigskip\medskip

Fabio Apruzzi$^a$, Jakob C. Geipel$^b$,\\ Andrea Legramandi$^{c}$,  Niall T. Macpherson$^{c,d}$ and Marco Zagermann$^e$ \\

\bigskip\medskip
{\small 
a: Department of Physics, University of North Carolina,\\ Chapel Hill, NC 27599, USA\\and \\Department of Physics and Astronomy,\\ University of Pennsylvania, Philadelphia, PA 19104, USA.\\[2mm]

b: Institut f\"ur Theoretische Physik, Leibniz Universit\"at Hannover\\
Appelstra\ss e 2, 30167 Hannover, Germany.\\[2mm]

c: Dipartimento di Fisica, Universit\`a di Milano--Bicocca, \\ Piazza della Scienza 3, I-20126 Milano, Italy \\ and \\ INFN, sezione di Milano--Bicocca.\\[2mm]

d: SISSA International School for Advanced Studies and INFN, sezione di Trieste,\\ 34136, Trieste, Italy.\\[2mm]

e: Fachbereich Physik der Universit\"{a}t Hamburg,\\
Luruper Chaussee 149, 22761 Hamburg, Germany.\\[2mm]
	
}

\bigskip\medskip

\vskip 0cm 
{\small \tt fabio.apruzzi@unc.edu, a.legramandi@campus.unimib.it, jakob.geipel@itp.uni-hannover.de, nmacpher@sissa.it, marco.zagermann@desy.de}
\vskip .9cm 
     	{\bf Abstract }

\vskip .1in
\end{center}

\noindent
We classify $\mathcal N=2$ Minkowski$_4$ solutions of IIB supergravity with an $SU(2)_R$ symmetry geometrically realized by an $S^2$-foliation in the remaining six dimensions. For the various cases of the classification, we reduce the supersymmetric system of equations to PDEs. These cases often accommodate systems of intersecting branes and half-maximally supersymmetric AdS$_{5,6,7}$ solutions when they exist. As an example, we analyze the AdS$_6$ case in more detail, reducing the supersymmetry equations to a single cylindrical Laplace equation. We also recover an already known linear dilaton background dual to the $(1,1)$ Little String Theory (LST) living on NS5-branes, and we find a new Minkowski$_5$ linear dilaton solution from brane intersections. Finally, we also discuss some simple Minkowski$_4$ solutions based on compact conformal Calabi-Yau manifolds.

\vfill
\eject

\end{titlepage}

\tableofcontents

\section{Introduction}
String compactifications to four-dimensional Minkowski spacetime (Minkowski$_4$) have traditionally enjoyed great interest as starting points for semi-realistic model-building in various approaches to string phenomenology. When the metric is the only nontrivial field of the solution, the presence of four Minkowski space-time dimensions implies that the rest of the real six-dimensional geometry, $M_6$, is Ricci-flat,  i.e. $T^6$, $T^2 \times K3$ or CY$_3$, or non-compact versions thereof, if some supersymmetry remains unbroken \cite{gibbons-nogo,dewit-smit-haridass,maldacena-nunez}.

Compactifications with nontrivial p-form fluxes, by contrast, are in general much harder to find and classify, since the presence of nontrivial NSNS and RR potentials modifies the Einstein equations, as well as the supersymmetry equations, such that these extra fields back-react on $M_6$. Except for some special cases (see e.g. \cite{grana-polchinski,giddings-kachru-polchinski}
and the related earlier works \cite{Dasgupta:1999ss,Gukov:1999ya,Becker:1996gj,Greene:2000gh}), where this back-reaction only introduces a conformal factor, the geometry of the internal space is in general drastically deformed away from Ricci-flatness and takes instead the form of a generalized Calabi-Yau manifold in the framework of generalized complex geometry \cite{gmpt2,hitchin-gcy,gualtieri,gmpt3}. Other successful attempts to construct explicit $M_6$ have been made in \cite{andriot,andriot-blaback-vanriet,andriot2,candelas-constantin-damian-larfors-morales-1,candelas-constantin-damian-larfors-morales-2}. 

As an additional difficulty, NSNS and RR potentials are magnetically and electrically associated to brane sources, which are localized on submanifolds of $M_6$ and likewise back-react on the geometry in a complicated way.
In fact, the difficulty of finding flux compactifications is in many ways related to the difficulty of finding intersecting brane solutions \cite{youm, imamura-D8,janssen-meessen-ortin,Bobev:2016phc}. An interesting role in this context is played by 
Anti-de-Sitter (AdS) compactifications  of string theory, which often arise as near-horizon limits of intersecting brane systems. Having the corresponding intersecting brane solutions at hand would be very useful to study holographic RG-flows of the dual field theories within the gauge/gravity correspondence, where the AdS vacuum would correspond to a conformal fixed point at one of the two ends of the RG-flow.

Another interesting corner of intersecting brane solutions is given by linear dilaton backgrounds, which can be seen as particular limit of these brane systems. These are conjectured to be dual to little string theories, which are very peculiar six-dimensional non-local theories with several string-like properties \cite{seiberg-lst,aharony-lst,kutasov-lst}. Since they are non-Lagrangian, holography represents a useful tool to study these models. 

In this paper, we systematically analyze supersymmetric Minkowski$_4$ $\times$ $S^2$ solutions of IIB supergravity, following the approach of \cite{Macpherson:2016xwk,Macpherson:2017mvu}. The geometry of $M_6$ is an $S^2$-foliation over a four-dimensional manifold, $M_4$, which can be compact or non-compact in order to accommodate also higher-dimensional Minkowski solutions as well as AdS$_{5,6,7}$, whenever these exist. In many cases $M_4$ admits an identity structure, i.e. there are enough spinors and bilinears to determine an entire vielbein.  The $S^2$ factor is sufficient to preserve at least $8$ real supercharges, and, in particular for $AdS$, the $S^2$-foliation geometrically implements the superconformal $SU(2)_R$ symmetry of the dual conformal field theory as part of the isometry of the IIB supergravity solution. So what is left to determine is the geometry of $M_4$ and how the other fields, such us fluxes, warp factors and the dilaton depend on its coordinates. 

The classification consists of three subclasses that correspond to: Mink$_6$ solutions, with supersymmetry conditions reduced to a single PDE, i.e. a 6D Laplace equation on $M_4$, which is just the PDE associated to D5-branes solutions; Mink$_5$ solutions, where supersymmetry is reduced to a system of three PDEs, and in general accommodates D5-NS5-D7-brane solutions; and finally Mink$_4$ solutions, which, in particular, encompasses all the conformal Calabi-Yau cases with an $S^2$-foliation.\footnote{Note that while a compact $CY_3$ manifold cannot have continuous isometries and also $T^6$ and $T^2\times K3$ do not admit SU(2) as an isometry group, there is no principal obstruction for  \emph{conformal} Calabi-Yau manifolds to have $SU(2)$ as an isometry group even when they are compact. In fact, the conformal factor may change the topology of the space and  compactify an originally non-compact Calabi-Yau space. A simple example is given by $S^2$, which is conformal to flat 2D space and admits $SU(2)$ as an isometry, whereas the compact and Ricci-flat CY analogue $T^2$ does not.}  In addition,  generalizations of  the first two classes are provided, which are governed by the same PDEs but have additional fluxes and a more complicated geometric structure.

As a second result, we also provide some explicit examples. We present a simple but new compact conformal Calabi-Yau example. Moreover, we recover all the possible AdS$_5$ solutions with an $S^2$-foliation, which are AdS$_5\times S^5$ and supersymmetric orbifolds thereof. We rewrite the IIB AdS$_6$ supersymmetric system of equations as a single cylindrical Laplace equation, and with an ansatz we generate infinitely many solutions. We also recover the result of \cite{Apruzzi:2013yva}, that there are no supersymmetric AdS$_7$ IIB solutions. Finally, we recover an already known linear dilaton background coming from NS5-branes, and some new example of linear dilaton background in Mink$_5$ with more involved brane intersections. 

The paper is organized as follows: in section \ref{sec:Ansatz}, we describe the field and spinor ans\"atze. In section \ref{sec:PurSpin}, we manipulate and reduce the pure spinor equations. In section \ref{sec: classification}, we exploit the classification, and in section \ref{sec:examples}, we give some explicit examples. The Appendix gives an infinite number of $AdS_6$ examples.

\section{$SU(2)_R$ preserving ansatz}\label{sec:Ansatz}
We are interested in finding $\mathcal{N}=2$ Mink$_4$ solutions of type II, so that their metric and RR flux poly-form may be expressed as
\beq\label{eq:Mink4ansatz}
d s^2_{10} = e^{2A} ds^2(\text{Mink}_4) + ds^2(M_6),~~~
F=F_{\text{int}} + e^{4A}\text{Vol}_4 \wedge \star_6 \lambda(F_{\text{int}}),
\eeq
where the dilaton $e^{\Phi}$, the NS 3-form $H$, the Minkowski warp factor $e^{2A}$ and the RR-forms $F_{\text{int}}$ are functions and forms on  $M_6$ only,
so that we preserve the $SO(1,3)$ isometry of $\textrm{Mink}_{4}$, and $\lambda(\omega_k) = (-1)^{k(k-1)/2} \omega_k$, for a k-form $\omega_k$.

Since we have extended supersymmetry we may have an R-symmetry. Motivated by the fact that it is a necessary part of the $\mathcal{N}=2$ super-conformal algebra in $d=4,5,6$, and  that it simplifies matters, we shall assume that we have an $SU(2)$ R-symmetry $SU(2)_R$. In general this should be realised geometrically as an $SU(2)$ isometry on $M_6$, which we shall specifically\footnote{This is not the only way to realise $SU(2)$, indeed it is possible to decompose $M_6$ as a fibration of $S^3$ over some $M_3$ in terms of the Maurer-Cartan of $SU(2)$. It is however unclear whether anything beyond the $SU(2)\times U(1)$ preserving squashed 3-sphere is compatible with $SU(2)_R$. When it is compatible one can always  T-dualise on the Hopf fibre and end up in a class with a round $S^2$ factor. Thus up to T-duality the combined results of this work and \cite{Macpherson:2016xwk} cover such cases} realise by ensuring that $M_6$ can be locally decomposed as $S^2\times M_4$ i.e.
\beq\label{eqS2M4decompostion}
ds^2(M_6) = e^{2C}ds^2(S^2) + ds^2(M_4),~~~ F_{\text{int}} = f +e^{2C} g\wedge \text{Vol}(S^2),~~~ H= H_3 + e^{2C} H_1 \wedge \text{Vol}(S^2)
\eeq
where $e^{2C},~ g,~ f, H_3,~H_1$ and the rest of the physical fields are functions or forms on $M_4$ only so as to preserve the $SU(2)$ isometry of the $S^2$ factor. 

The most general Majorana-Weyl (MW) Killing spinors in 10 dimensions that are consistent with a Mink$_4\times S^2\times M_4$ decomposition  \cite{Macpherson:2016xwk}  in IIB are of the form
\begin{equation}\label{eq 10dspinors}
\epsilon_{1} = \sum_{b =1}^2 \zeta_+^b \otimes \xi^b_+ \otimes \eta^{1}_+ +  \zeta_+^b \otimes \xi^b_- \otimes \eta^{1}_- + \text{m.c.},~~~\epsilon_{2} = \sum_{b =1}^2 \zeta_+^b \otimes \xi^b_+ \otimes \eta^{2}_+ +  \zeta_+^b \otimes \xi^b_- \otimes \eta^{2}_- + \text{m.c.},
\end{equation}
where $\zeta_+^b$ is a doublet of spinors on Mink$_4$, $\xi^{b}_{\pm}$ are doublets of Killing spinors on $S^2$, $\eta^{i}_{\pm}$ are arbitrary spinors on $M_4$, $\pm$ labels chirality, and m.c. stands for Majorana conjugate. The doublets on $S^2$ are of the form
\beq
\xi^{b}= \left(\begin{array}{c} \xi \\ \xi^c\end{array}\right),
\eeq
where $\xi=\xi_++\xi_-$ and $\xi^c= \sigma_2 \xi^{*}$, for $\sigma_i$ the Pauli matrices,  is the Majorana conjugate of $\xi$. This doublet is charged under $SU(2)$ as can be seen if one calculates the spinoral Lie derivative along the $SU(2)$ Killing vectors $K_i$:
\beq\label{su2transformation}
\mathcal{L}_{K_i}\xi^{b} = \frac{i}{2} (\sigma_i)^b_{~c} \xi^c
\eeq
which realised the algebra of $SU(2)$ and is how $SU(2)_R$ is realised at the level of spinors. The appealing thing about this set-up is on the one hand
 that whenever we perform a global $SU(2)$ transformation on $\xi^b$, we can perform a simultaneous transformation on $\zeta^a_{+}$ such that the 10 dimensional spinors in \eqref{eq 10dspinors} remain invariant - much like realising a global symmetry with a Lagrangian in field theory. On the other hand, \eqref{su2transformation} means that $\xi^\b$ is charged under local $SU(2)$ transformation. Under such a local transformation the parts of \eqref{eq 10dspinors} that couple to $\zeta^1$ and $\zeta^2$ are mapped to each other, which means that if we solve one part in terms of conditions on $M_4$ then the other part is also guaranteed - this allows us to focus on an $\mathcal{N}=1$ sub-sector only. This reduces our consideration to 6D Killing spinors of the form
\beq\label{eq: 6dspinorsansatz}
\chi^1_+= \xi_+\otimes \eta^1_++\xi_-\otimes \eta^1_-,~~~~\chi^2_+= \xi_+\otimes \eta^2_++\xi_-\otimes \eta^2_-.
\eeq
The norms of these 6D spinors are fixed to be proportional to the warp factor $e^A$ \cite{gmpt3}, while the norms of the chiral spinors on $S^2$ are charged under $SU(2)_R$ such that only $|\xi_+|^2+ |\xi_-|^2$ is a singlet. This means that our assumption that $e^{2A}$ respects the $SU(2)$ isometry leads to the following restriction on the 4D spinor norms
\begin{equation}\label{eq:4dansatz1}
|| \eta^a_+ ||^2 = || \eta^a_-||^2 ,~~~a=1,2.
\end{equation}
This is all that we require on isometry grounds, but we also choose to impose
\begin{equation}\label{eq:4dansatz2}
|| \eta^1_+ ||^2 = || \eta^2_-||^2.
\end{equation}
This is a reasonable simplifying assumption which implies that the six-dimensional spinors $\chi^1$ and $\chi^2$ have equal norm. This is required globally for $AdS_{d}$ solutions with $d=4,5,6,7$ and is a local requirement for the existence of calibrated D-branes and O-planes, but is not needed in general.\\

Having explained the specifics of our spinor and geometrical approach, we will now proceed to reduce the problem of finding solutions to geometric constraints on $M_4$ only.
\section{Pure spinor conditions in 4D}\label{sec:PurSpin}
We have established that we only need to solve an $\mathcal{N}=1$ sub-sector of  the full $\mathcal{N}=2$ spinors to find a solution, and given that $\mathcal{N}=1$ solutions with warped Mink$_4$ factors were classified in \cite{gmpt2}, it seems sensible to use this as a starting point. The conditions of unbroken supersymmetry in IIB are equivalent to the existence of two 6D pure spinors
\beq
\Phi_+= e^{-A}\chi^1\otimes \chi^{2\dag},~~~~ \Phi_- = e^{-A}\chi^1\otimes \overline{\chi}^{2},~~~~\overline{\chi}^{2}= ((\chi^{2})^c)^{\dag}
\eeq
 which for us, given \eqref{eq:4dansatz2}, must satisfy
\begin{subequations}
\label{6dSUSYconds}
\begin{align}
&|\chi^1|^2= |\chi^2|^2 =  e^{A},\label{6dSUSYconds a}\\[2mm]
&d_H(e^{3A-\Phi}\Phi_-)=0,\label{6dSUSYconds b}\\[2mm]
&d_H(e^{2A-\Phi}\text{Re}\Phi_+)=0,\label{6dSUSYconds c}\\[2mm]
&d_H(e^{4A-\Phi}\text{Im}\Phi_+)=\frac{1}{8}e^{4A}\star_6 \lambda(F_{\text{int}}),\label{6dSUSYconds d}
\end{align}
\end{subequations}
 According to the previous section, we take the 2D spinors to be
\beq
\chi^1_+= \frac{1}{2}e^{\frac{A}{2}}(\xi\otimes \eta^1+\sigma_3\xi\otimes \hat\gamma\eta^1),~~~\chi^2_+=\frac{1}{2}e^{\frac{A}{2}}(\xi\otimes \eta^2+\sigma_3\xi\otimes \hat\gamma\eta^2),
\eeq
which is just a rewriting of \eqref{eq: 6dspinorsansatz}, where the 4D spinors decompose as $\eta^i = \eta^i_++ \eta^i_-$, and similarly for $\xi$, with $\sigma_3$ and $\hat\gamma$ the 2D and 4D chirality matrices, respectively. Our first task is to decompose the 6D pure spinors as wedge products of bi-spinors on $S^2$ and $M_4$, this is easily done by making repeated use of the identity\footnote{It is best to derive this with respect to a specific representation of the flat space 6D gamma matrices, here and elsewhere we use
\beq
\gamma^{(6)}_{a}=  \sigma_a\otimes \mathbb{I}_4,~~~ \gamma^{(6)}_{i+2}= \sigma_3 \otimes \gamma_i,~~~B_6 = \sigma_2\otimes B_4,
\eeq
where $a=1,2$ and $i=1,...,4$, $\sigma_a$ are the Pauli matrices. The 4D $\gamma_i$ are such that
\beq
\gamma_i^*=B_4^{-1} \gamma_i B_4
\eeq
and the 4d intertwiner must be such that $B_4 B_4^{*}= -\mathbb{I}_4$. }
\beq
(\xi^1\otimes \eta^1)\otimes (\xi^2\otimes \eta^2)^{\dag}= \xi^1\otimes\xi^{2\dag}\wedge (\eta^2\otimes \eta^{2\dag})_+ + (\sigma_3\xi^1\otimes\xi^{2\dag})_+\wedge (\eta^2\otimes \eta^{2\dag})_-- (\sigma_3\xi^1\otimes\xi^{2\dag})_-\wedge (\eta^2\otimes \eta^{2\dag})_-,\nn
\eeq
where $\pm$ now refers to the even/odd form part of a given bilinear. The bi-spinors on an $S^2$ of radius $e^{2C}$ can be parameterised as 
\begin{subequations}\label{eq:bispinorsS2}
\begin{align}
(\xi\otimes \xi^{\dag})_+&=\frac{1}{2}(1- i y_3 e^{2C}\text{Vol}(S^2)), \quad\quad~ (\xi\otimes \xi^{\dag})_- =\frac{1}{2} e^{C} K_3,\\
(\sigma_3 \xi\otimes \xi^{\dag})_+&=\frac{1}{2}(y_3- i e^{2C}\text{Vol}(S^2)),\quad\quad (\sigma_3 \xi\otimes \xi^{\dag})_- = \frac{i}{2} e^{C} dy_3,\\
(\xi\otimes \overline{\xi})_+&=-\frac{i}{2}(y_1+i y_2) e^{2C}\text{Vol}(S^2),\quad\quad(\xi\otimes \overline{\xi})_- =\frac{1}{2} e^{C}(K_1+ i K_2) ,\\
(\sigma_3\xi\otimes \overline{\xi})_+&=\frac{1}{2}(y_1+i y_2),\quad\quad\quad\quad\quad\quad (\sigma_3\xi\otimes \overline{\xi})_- =\frac{i}{2} e^C(dy_1+ i dy_2),
\end{align}
\end{subequations}
where $K_i$ are $SU(2)$ Killing vectors and $y_i$ are coordinates embedding $S^2$ into $\mathbb{R}^3$ \cite{Macpherson:2016xwk}. The key point to recognise here is that everything appearing in \eqref{eq:bispinorsS2} is part of a closed set of forms under the action of $d$ and $\wedge$ (note $dK_i=2y_i \text{Vol}(S^2)$) - this allows us to reduce \eqref{6dSUSYconds a}-\eqref{6dSUSYconds d} to a set of pure-spinor relations in 4D given the flux decomposion in \eqref{eqS2M4decompostion}. First one finds that the 6D pure spinors may be expressed as
\begin{align}
\Phi_- =& \frac{1}{4} \Big[ (y_1+ i y_2) \Psi^2_- + e^C (K_1+i K_2)\wedge  \Psi^2_+
+ i \Big( e^C d (y_1+ i y_2) \wedge \Psi^2_{\hat\gamma_+}  - (y_1+ i y_2)   e^{2C}\text{Vol}(S^2)\wedge \Psi^2_{\hat\gamma_-} \Big)   \Big]\nn\\[2mm]
\Phi_+=& \frac{1}{4} \Big[\Psi^1_+ + y_3\Psi^1_{\hat\gamma+}  - e^C K_3 \wedge\Psi^1_{\hat\gamma-}
- i \Big( e^C d y_3 \wedge\Psi^1_{-}   + e^{2C} \text{Vol}(S^2) \wedge \Psi^1_{\hat\gamma+}    + e^{2C} y_3 \text{Vol}(S^2) \wedge\Psi^1_{+} \Big) \Big]\nn,
\end{align}
where we have defined the following 4D pure spinors
\begin{align}\label{eq:4dpurespinors}
\Psi^1&= \eta^1\otimes \eta^{2\dag},~~~\Psi_{\hat\gamma}^1= (\hat\gamma\eta^1)\otimes \eta^{2\dag},\\[2mm]
\Psi^2&= \eta^1\otimes \overline{\eta}^{2},~~~\Psi_{\hat\gamma}^2= (\hat\gamma\eta^1)\otimes \overline{\eta}^{2}.
\end{align}
Upon plugging this back into the 6D supersymmetry conditions one indeed finds that each of them is implied by pure spinor conditions on $M_4$. Specifically, one finds two independent complex constraints from \eqref{6dSUSYconds b}
\begin{subequations}
\begin{align}
&d_{H_3}\big(e^{3A+C-\Phi}\Psi^2_{\hat\gamma+}\big)+ i e^{3A-\Phi}\Psi^2_-=0,\label{eq:4dsusy1a}\\[2mm]
&d_{H_3}\big(e^{3A+2C-\Phi}\Psi^2_{\hat\gamma-}\big)- i e^{3A+2C-\Phi}H_1\wedge \Psi^2_-+ 2 i e^{3A+C-\Phi}\Psi^2_+=0,\label{eq:4dsusy1b}
\end{align}
\end{subequations}
four real constraints from \eqref{6dSUSYconds c} 
\begin{subequations}
\begin{align}
&d_{H_3}\big(e^{2A-\Phi}\text{Re}\Psi^1_+\big)=0, \label{eq:4dsusy2a}\\[2mm]
&d_{H_3}\big(e^{2A+C-\Phi}\text{Im}\Psi^1_-\big)-e^{2A-\Phi}\text{Re}\Psi^1_{\hat\gamma+}=0, \label{eq:4dsusy2b}\\[2mm]
&d_{H_3}\big(e^{2A+2C-\Phi}\text{Im}\Psi^1_{\hat\gamma+}\big)- e^{2A+2C-\Phi}H_1\wedge \text{Re} \Psi^1_+=0, \label{eq:4dsusy2c}\\[2mm]
&d_{H_3}\big(e^{2A+2C-\Phi}\text{Im}\Psi^1_{+}\big)- e^{2A+2C-\Phi}H_1\wedge \text{Re}\Psi^1_{\hat\gamma+}- 2 e^{2A+C-\Phi}\text{Re}\Psi^1_{\hat\gamma_-}=0, \label{eq:4dsusy2d}
\end{align}
\end{subequations}
two  real constraints from \eqref{6dSUSYconds d}
\begin{subequations}
\begin{align}
&d_{H_3}\big(e^{4A+C-\Phi}\text{Re}\Psi^1_-\big)+ e^{4A-\Phi}\text{Im}\Psi^1_{\hat\gamma+}=0,\label{eq:4dsusy3a}\\[2mm]
&d_{H_3}\big(e^{4A+2C-\Phi}\text{Re}\Psi^1_+\big)+e^{4A+2C-\Phi}H_1\wedge \text{Im}\Psi^1_{\hat\gamma+}+ 2e^{4A+C-\Phi}\text{Im}\Psi^1_{\hat\gamma-}=0,\label{eq:4dsusy3b}
\end{align}
\end{subequations}
as well as the definition of the fluxes through
\begin{subequations}
\label{RR_fluxes}
\begin{align}
&d_{H_3}\big(e^{4A-\Phi}\text{Im}\Psi^1_+\big)+ \frac{1}{2} e^{4A}\star_4 \lambda(g)=0,\label{eq:4dfluxa}\\[2mm]
&d_{H_3}\big(e^{4A+2C-\Phi}\text{Re}\Psi^1_{\hat\gamma+}\big)+ e^{4A+2C-\Phi}H_1\wedge \text{Im}\Psi^1_++\frac{1}{2} e^{4A+2C}\star_4\lambda(f)=0 \label{eq:4dfluxb}.
\end{align}
\end{subequations}
At first glance, this looks like a rather complicated system of form equations, but they are actually rather restrictive. Indeed,  it is already clear that we must have the following zero form constraints
\beq\label{eq:zeroform1}
\Psi^2_0= \Psi^1_{\hat\gamma 0}=0,
\eeq
which follow from \eqref{eq:4dsusy1b}, \eqref{eq:4dsusy2b} and \eqref{eq:4dsusy3a}. The condition \eqref{6dSUSYconds a} also furnishes us with additional constraints
\beq\label{eq:zeroform2}
|\eta^1|^2=|\eta^2|^2= 1,~~~\eta^{1\dag}\hat\gamma\eta^1=\eta^{2\dag}\hat\gamma\eta^2=0.
\eeq

In order to make further progress, we must parameterise the 4D pure spinors \eqref{eq:4dpurespinors}. In general, we can expand any non-chiral 4D spinor in a basis of $\{\eta,\hat\gamma \eta,\eta^c,\hat\gamma \eta^c\}$, and so we decompose $\eta^2$ in a basis of $\eta^1$ in this fashion. We must also solve the zero form conditions, however, which kill some of the possible terms. The most general decomposition  consistent with \eqref{eq:zeroform1}-\eqref{eq:zeroform2} is
\beq
\eta^1= \eta,~~~ \eta^2= a \eta+ b \hat\gamma \eta^c, ~~~|\eta|^2=1,~~~\eta^{\dag}\hat\gamma\eta=0,\label{eq: etadecompostion}
\eeq
where $a$ and $b$ are arbitrary complex functions on $M_4$ subject to the constraint
\beq
|a|^2+|b|^2= 1.
\eeq
The final ingredient one needs to write down the pure spinors is that unit norm chiral spinors define a vielbein on $M_4$, namely
\beq
v= v_1+ i v_2=\eta_-^{\dag}\gamma_a\eta_+ dx^a,~~~ w= w_1+ i w_2=\overline{\eta}_-\gamma_a\eta_+dx^a
\eeq
which was derived in \cite{Apruzzi:2014qva}. This is already sufficient to refine \eqref{eq: etadecompostion}: Notice that \eqref{eq:4dsusy3b} gives rise to the 1-form condition
\beq
d(e^{3A+C-\Phi} b)+ b e^{3A-\Phi} \text{Im}v=0,
\eeq
which is sufficient to establish that for $b= |b|e^{i \text{Arg}(b)}$ we must have $d(\text{Arg}(b))=0$.
This leads us to parameterise 
\beq
a= e^{-i\zeta}\kappa_{\|},~~~~ b= e^{i\zeta_0} \kappa_{\perp},~~~~ d\zeta_0=0,~~~ \kappa_{\perp}^2+ \kappa_{\|}^2=1,
\eeq
and perform the simultaneous rotations of the vielbein and $S^3$ embedding coordinates
\beq 
w\to e^{i (\zeta+\zeta_0)} w,~~~~ y_1+i y_2 \to e^{-i \zeta_0}(y_1+ i y_2).
\eeq
These make $\Phi_{\pm}$ independent of $\zeta_0$ and make $\zeta$ appear only in $\Phi_+$ as an overall phase. We are now ready to calculate the pure spinors of \eqref{eq:4dpurespinors}. We find that their even and odd form components are expressed in terms of the vielbein on $M_4$ as
\begin{subequations}\label{eq: 4dpurespinors}
\begin{align}
\Psi^1_+&= \frac{1}{2}e^{i\zeta}\kappa_{\|} e^{\frac{1}{2}w\wedge \overline{w}- \frac{\kappa_{\perp}}{\kappa_{\|}} v_1\wedge w},~~~\Psi^2_+=- \frac{i}{2} v_2\wedge(\kappa_{\perp} v_1+ \kappa_{\|} w)\wedge e^{\frac{1}{2} w\wedge \overline{w}},\\[2mm]
\Psi^1_{\hat\gamma+}&=\frac{i}{2} e^{i\zeta} v_2\wedge(\kappa_{\|} v_1- \kappa_{\perp}w) \wedge e^{\frac{1}{2}w \wedge \overline{w}},~~~\Psi^2_{\hat\gamma+}= - \frac{1}{2} \kappa_{\perp} e^{\frac{1}{2} w \wedge \overline{w}+ \frac{\kappa_{\|}}{\kappa_{\perp}}v_1\wedge w},\\[2mm]
\Psi^1_- &= \frac{1}{2}e^{i\zeta} (\kappa_{\|} v_1-\kappa_{\perp} w)\wedge e^{\frac{1}{2}w \wedge \overline{w}},~~~ \Psi^2_-= \frac{i}{2} \kappa_{\perp} v_2 \wedge e^{\frac{1}{2}w\wedge \overline{w}+ \frac{\kappa_{\|}}{\kappa_{\perp}} v_1\wedge w},\\[2mm]
\Psi^1_{\hat\gamma -}&= \frac{i}{2} e^{i \zeta} \kappa_{\|} v_2 e^{\frac{1}{2}w \wedge \overline{w}- \frac{\kappa_{\perp}}{\kappa_{\|}}v_1\wedge w},~~~ \Psi^2_{\hat\gamma-}= \frac{1}{2} (\kappa_{\perp} v_1+ \kappa_{\|} w)\wedge e^{\frac{1}{2}w\wedge \overline{w}}.
\end{align}
\end{subequations}
This parametrisation will be key to work out all the classes that follow from our broad $SU(2)_R$ preserving spinor ansatz.
Note that the 1-form part of \eqref{eq:4dsusy2c} imposes that 
\beq\label{eq: physicalcondtion}
\cos\zeta \kappa_{\|}H_1=0,
\eeq
which means at least one of these must always be zero and  already hints at the branching off of different  classes of solution in the following section.

Before we press on with the classification let us look at how any potential solution will fall within the known classification of $SU(3)\times SU(3)$-structures in 6d.  Given everything we have established in this section, it is possible to show that the 6d pure spinors are of the form
\begin{subequations}\label{eq: 6dbispinors}
\begin{align}
\Phi_+&=\frac{e^{i\zeta}}{8}  e^{\frac{1}{2}E_3\wedge \overline{E}_3}\wedge\bigg(\kappa_{\|} e^{\frac{1}{2}(E_1\wedge \overline{E}_1+E_2\wedge \overline{E}_2)}+ i \kappa_{\perp}E_1\wedge E_2\bigg) ,\\[2mm]
\Phi_-&= \frac{i}{8} E_3\wedge\bigg(\kappa_{\perp}e^{\frac{1}{2}(E_1\wedge \overline{E}_1+E_2\wedge \overline{E}_2)}- i \kappa_{\|} E_1\wedge E_2\bigg)
\end{align}
\end{subequations}
i.e. an intermediate 	\cite{Andriot:2008va}, possibly dynamical \cite{Macpherson:2013zba,Gaillard:2013vsa,Andriot:2014qla}, $SU(2)$-structure when $\kappa_{\perp}\neq 0$, an $SU(3)$-structure when  $\kappa_{\perp}=0$ so that the 6d spinors are strictly parallel, or an orthogonal $SU(2)$-structure when  $\kappa_{\|}=0$ and the 6d spinors are strickly orthogonal. The 6d complex vielbein is
\beq
\label{eq: 6dcomplexvielbein}
E_1= e^{C}dy_3- y_3 v_2+ i v_1,~~~ E_2= w,~~~ E_3= - ( e^{C}(dy_1+ i dy_2)- (y_1+ i y_2) v_2).
\eeq
The 6D bilinears in the form \eqref{eq: 6dbispinors} also give us useful information about  what branes can be embedded in a solution. In fact the supersymmetry conditions that we get from $\kappa$-symmetry can be rephrased in terms of the two bilinears $\Phi_{\pm}$, which can be viewed as calibrations for the internal cycle the brane wraps \cite{Martucci:2005}.
In the case of four-dimensional vacuum solutions in IIB supergravity the calibrations for space-time filling branes are given by $e^{4A-\Phi} \text{Im} \Phi_+ $. The parts of this bilinear compatible with the $SU(2)$ isometry, for the cases we will see in the classification, are:
\begin{equation}
\begin{split}
	&\text{Case I:} \qquad
 -e^{4A-\Phi} w_1 \wedge w_2 +e^{4A+2C-\Phi} \text{Vol}(S^2)\wedge v_1 \wedge v_2 \wedge w_1 \wedge w_2  \\
	&\text{Case II:} \qquad
	-e^{4A-\Phi} v_1 \wedge w_1 + 
 e^{4A+2C-\Phi} \text{Vol}(S^2)\wedge 
	v_2 \wedge w_1 \\
	&\text{Case III:} \qquad e^{4A-\Phi} -  e^{4A+2C-\Phi} \text{Vol}(S^2)\wedge v_1 \wedge v_2 
\end{split}
\end{equation}
where Case I,II and III refers respectively to \ref{sec:D5}, \ref{sec:mink5system} and \ref{sec:conformalCalab} of the next section.

In the next section, we shall find local expressions for all solutions to \eqref{eq:4dsusy1a}-\eqref{eq:4dfluxb} and the Bianchi identities of the fluxes up to PDEs.  As we shall see the physical interpretations of the various cases that follow will be quite different.
\section{Classification}\label{sec: classification}
In this section we shall classify every type of solution that follows from the ans\"atze of section \ref{sec:Ansatz} - this means finding every \eqref{eq: 4dpurespinors} that solves \eqref{eq:4dsusy1a}-\eqref{eq:4dsusy3b}. In \eqref{eq: 4dpurespinors}, we have both $\kappa_{\perp}$ and $\kappa_{\|}$ appearing in the denominators of certain bi-spinors so that we need to look at the cases where these vanish individually before examining more generic situations. As we shall see, it is \eqref{eq: physicalcondtion} that really determines the ultimate physical interpretation of the various cases we find, namely which of $\cos\zeta$, $k_{\|}$ and $H_1$ is set to zero. In all cases, we shall be able to give the explicit local form the metric and fluxes must take up to PDEs. To be more precise, we mean the local form in the sense that all coordinate patches of a global Mink$_4\times S^2$ solution can be expressible in this form.

Solving \eqref{eq:4dsusy1a}-\eqref{eq:4dsusy3b} is not by itself sufficient for a solution to exist - we also need to solve the Bianchi identities of the fluxes. Given a set of 4D pure spinors,  \eqref{RR_fluxes} tells us what RR flux necessarily follows, however this only explicitly tells us the part of the flux parallel to Mink$_4$ - the electric part. To get the internal magnetic components of the flux we are required to take the Hodge dual on $M_4$ - doing this for an arbitrary undetermined metric is a highly nontrival task, although technology does exist to aid the process \cite{Tomasiello:2007zq}. However, as we shall be able to define the local metric on $M_4\times S^2$, this step will be comparatively easy for us. 

The Bianchi identities (BIs) of the electric component of the fluxes are actually implied by supersymmetry, as should be evident from  \eqref{eq:4dfluxb} given \eqref{eq:Mink4ansatz}, so it is only the magnetic flux BIs that we must solve. Away from localised sources they take the form $d_H F_{\text{int}\textit{}}=0$, which in terms of \eqref{eqS2M4decompostion} become
\beq\label{eq: 4d Bianshis}
d_{H_3}f= d_{H_3}(e^{2C}g)- e^{2C}H_1\wedge f=0.
\eeq
Solving these and \eqref{eq:4dsusy1a}-\eqref{eq:4dsusy3b} is sufficient for a solution to exist \cite{Koerber:2007hd} everywhere, except at the loci of brane-like (i.e. $\delta$-function dependent) sources, and gives rise to the PDEs discussed in this section. 
We would like to stress, however, that this does not mean that solutions to the PDEs are incompatible with localised sources (indeed they may necessarily follow), merely that some extra care needs to be taken when they are present.

More concretely, one can describe the  process of finding solutions as follows: One first finds a solution to the PDEs following from supersymmetry and \eqref{eq: 4d Bianshis} that we present in this section - this defines fluxes and geometry. One then checks whether the metric and/or dilaton are signalling a singularity anywhere in this geometry - if the solution is regular one is done. Any singularity that is present should have a physical string-theory origin, such as (but not limited to) a D-brane or O-plane and so the metric should exhibit a behaviour consistent with this - one should only proceed further if this is at all plausible (see e.g. 
\cite{Blaback:2011pn,Gautason:2013zw,Apruzzi:2013yva} for some concrete examples and further details). Having established a plausible brane-like scenario for a singularity one should then check whether a) this is compatible with supersymmetry, by checking the $\kappa$-symmetry conditions or where appropriate the calibrations \cite{Koerber:2007hd}, and whether b)  this obeys the Bianchi identities at the local source, which is the only place one ``sees'' the delta function. However, from a practical perspective, one often finds that close to a singularity the metric and fluxes reproduce a known brane behaviour asymptotically - when this happens one already knows that a) and b) will be satisfied.\\

We can now proceed with  the classification:

\subsection{Case I: The D5-brane - Mink$_6$}\label{sec:D5}
The first thing we need to do is choose a way to solve \eqref{eq: physicalcondtion} - for this case we will set $H_1=0$. We also choose to fix $(\kappa_{\|}=1,~\kappa_{\perp}=0,~\sin\zeta=0)$ and defer looking at generic values away from this choice until section \ref{sec: Gen I}. As we shall see, this case contains only the D5-brane - however it  does  give an example simple enough to be very detailed in our derivation - the methods we use will apply to all cases, but here we can be explicit without the discussion becoming protracted.

For this case, it is possible to show that \eqref{eq:4dsusy1a}-\eqref{eq:4dsusy3b} are implied by the necessary and sufficient conditions for supersymmetry
\begin{subequations}
 \label{SusyCase1B0}
\begin{align}
&d(e^{A+C})+ e^{A}v_2= d(e^{A}v_1)= d(e^{A-\Phi} w)=0,\label{SusyEq2Case1B0}\\[2mm]
&H_3=H_1=0=d(e^{2A-\Phi})=d(e^{2A})\wedge v_1\wedge v_2=0. \label{SusyEq3Case1B0}
\end{align}
\end{subequations}
As will be typical of all the cases we encounter, we can solve \eqref{vielbein_kperp_z=0} by
introducing local coordinates $x_1,x_3,x_4$ and
\beq
x_2= e^{A + C}
\eeq
that span $M_4$ and then solve \eqref{SusyEq2Case1B0} by using it to define the vielbein in terms of these local coordinates as
\begin{equation}
\label{vielbeinD5case}
v_1 = e^{-A} d x_1,~~~ v_2 = -e^{-A} d x_2, ~~~ w = e^{-A+\Phi}( d x_3+i dx_4).
\end{equation}
With these definitions it is then not hard to see that  \eqref{SusyEq3Case1B0} impose that there is no NS 3-form flux and that
\beq
e^{-4A}=e^{-2\Phi}=: h(x_1,x_2),
\eeq
up to rescaling $g_s$. This implies that  $\partial_{x_3}$ and $\partial_{x_4}$ are isometry directions, which can be taken to span either $T^2$ or $\mathbb{R}^2$, which are locally the same. As the warp factor is $e^{2A}$, we choose to take the latter possibility so that Mink$_4~\to~$Mink$_6$. The 10d metric is then given by
\begin{equation}
d s^2 =\frac{1}{\sqrt{h}}ds^2(\text{Mink}_6)+  \sqrt{h}\bigg(d x_1^2+d x_2^2+ x_2^2 d s^2(S^2) \bigg),
\end{equation}
which has the Minkowski factor and warping indicative of a D5-brane - but to confirm this we  need to derive the fluxes.

The flux component orthogonal to $S^2$, $f$, is defined by \eqref{eq:4dfluxa}, which reduces in this case to
\begin{equation}
\star_4 \lambda(f) = -2 e^{-4A-2C} d\left( e^{4A+2C-\Phi} \text{Re} \Psi^1_{\hat\gamma+} \right)
\end{equation}
given that $H_1=H_3=0$.  The pure spinor \text{Re} $\Psi^1_{\hat\gamma+}$ is defined in  \eqref{eq: 4dpurespinors}, which becomes
\begin{equation}
\text{Re} \Psi^1_{\hat\gamma+} = - \frac{1}{2} v_1 \wedge v_2 \wedge w_1 \wedge w_2 = - \frac{1}{2} \text{Vol}(M_4) 
\end{equation}
after we fix $(\kappa_{\|}=1,~\kappa_{\perp}=0,~\sin\zeta=0)$ - this means that $\star_4 \lambda(f)$ is proportional to the exterior derivative of the top form on $M_4$,
 which implies $f = 0$. 
The flux component parallel to $S^2$ is parameterised by $g$, its derivation is analogous: from  \eqref{eq: 4dpurespinors} we read off
\begin{equation}
\text{Im} \Psi^1_{+} = - \frac{1}{2}  w_1 \wedge w_2 = - \frac{1}{2} e^{2A} d x_3 \wedge d x_4
\end{equation} 
while \eqref{eq:4dfluxb} gives us the definition of $\star_4 g$
\begin{equation}
\star_4 \lambda(g) = -2 e^{-4A} d_{H_3} \left( e^{4A-\Phi} \text{Im} \Psi^1_{+} \right) = - e^{3A} \left( \partial_{x_1}e^{-4A} v_1 -  \partial_{x_2}e^{-4A} v_2  \right) \wedge w_1 \wedge w_2.
\end{equation} 
We can then use the local vielbein of  \eqref{vielbeinD5case} to take the Hodge dual of this expression and arrive at
\begin{equation}
g = e^{2A} \left( \partial_{x_2}e^{-4A} d x_1 - \partial_{x_1}e^{-4A} d x_2 \right) 
\end{equation}
which must be inserted in the definition of $F_{\text{int}}$ \eqref{eqS2M4decompostion} - note that $e^{2A}= x_2^2 e^{-2C}$.
So we find that the only non trivial 10  dimensional flux in this case is the RR 3-form
\beq
F_3 =  x_2^2\bigg(\partial_{x_2}hdx_1- \partial_{x_1}h dx_2\bigg)\wedge \text{Vol} (S^2).
\eeq

It is not hard to see that imposing its closure leads to
\beq\label{eq:D5Laplacian}
\frac{1}{x_2^2}\partial_{x_2}(x_2^2 \partial_{x_2}h)+ \partial_{x_1}^2h =0,
\eeq
which is 4D Laplace equation expressed in cylindrical polar coordinates. Indeed this is just the PDE associated to a D5-brane with $SO(3)$ symmetry due to the $S^2$ factor of our ansatz - one attains the more standard warped $S^3$ result by expressing $(x_1,~x_2)$ polar coordinates and then imposing that $h$ is independent of the angle. We shall find another class of solution in section \ref{sec: Gen I} that is also governed by \eqref{eq:D5Laplacian}, but which is incompatible with a Mink$_6$ factor. This case then exhausts Mink$_6\times S^2\times M_2$ solution in IIB (with equal internal spinor norm) and so gives every solution to (4.11) of \cite{Lust:2010by} with an $S^2$ factor.

\subsection{Case II: D7-D5-NS5-brane system - Mink$_5$} \label{sec:mink5system}
For this case we solve  \eqref{eq: physicalcondtion} with $\kappa_{\|}=0$, which fixes $\kappa_{\perp}=1$ without loss of generality - this means that the 6d pure spinors are those of an orthogonal $SU(2)$ structure. If one examines \eqref{eq: 4dpurespinors} in this limit, it should be clear that it is possible to fix the phase $\zeta$ to any value by rotating $(w_1,~w_2)$ - we choose $\zeta=\frac{\pi}{2}$ to make contact with the class of section \ref{sec: Gen II}.

For this case one finds that the supersymmetry conditions  \eqref{eq:4dsusy1a}-\eqref{eq:4dsusy3b} are implied by the constraints
\begin{subequations} \label{eq:Susy_5d}
\begin{align} 
&d(e^{3A+C-\Phi})+ e^{3A-\Phi}v_2=d(e^{A} w_2)=d(e^{-A} w_1)= d(e^{-3A+\Phi}(v_1+ B_0 v_2))=0,  \label{eq:Susy1_5d}\\[2mm]
&d(e^{A-\Phi}v_1)\wedge w_2= d(e^{A+2C-\Phi}(B_0 v_1- v_2))\wedge w_2=B_2=0, \label{eq:Susy2_5d}
\end{align}
\end{subequations}
where we have introduced potentials for the NS 3-form, $B_2$ and $B_0$,  such that
\beq\label{eq:LNSflux potential}
H=dB=d(B_2+ e^{2C} B_0 \text{Vol}(S^2)) =H_3+ e^{2C}H_1 \wedge \text{Vol}(S^2).
\eeq
As in the previous section, we can solve a good deal of these constraints by introducing local coordinates on $M_4$: $x_1,~x_3,~x_4$ and, this time
\begin{equation}
x_2 = e^{3A+C-\Phi}.
\end{equation}
In terms of these we can use \eqref{eq:Susy1_5d} to define a vielbein on $M_4$ as
\begin{equation}
v_1 = e^{3A - \Phi} d x_1 + B_0e^{-3A + \Phi} d x_2 ,~~~ v_2 = - e^{-3A + \Phi} d x_2 , ~~~ w_1 = e^{A} d x_4 , ~~~ w_2 = e^{-A} d x_3.
\end{equation}
On the other hand, \eqref{eq:Susy2_5d} tells us that the NS 3-from must be strictly parallel to $S^2$, that $B_0,~e^{A},~e^{\Phi}$ must satisfy the following PDEs
\begin{subequations}\label{3.2_BPS}
\begin{align}
&\partial_{x_2} (e^{4A - 2 \Phi}) = \partial_{x_1} (e^{-2A} B_0) , \label{3.2_BPS1} \\[2mm]
&\frac{1}{x^2_2} \partial_{x_2} (x_2^2 e^{-2A} B_0) = \partial_{x_1} (e^{-8A+2 \Phi}(1+B_0^2 ))  \label{3.2_BPS2},
\end{align}
\end{subequations}
and that $\partial_{x_4}$ is symmetry of the solution, parameterising either $S^1$ or $\mathbb{R}$ . We once  more observe that the isometry direction has the same warping as the Minkowski factor which locally gives an enhancement to Mink$_5$. The 10 dimensional metric is then of the form
\beq\label{eq: case II metric}
d s^2 =e^{2A}ds^2(\text{Mink}_5)+   e^{-6A + 2 \Phi} \bigg( d x_2^2+ x_2^2 d s^2 (S^2)  \bigg) + e^{-2A} d x_3^2+ e^{6A - 2 \Phi} \bigg(d x_1 + B_0 e^{-6A + 2 \Phi} d x_2\bigg)^2 .
\eeq
The 10 dimensional fluxes can now be extracted from \eqref{RR_fluxes} by following the prescription explained at length in section \ref{sec:D5}, which yields in the case at hand
\begin{align}
B&= e^{-6A+2\Phi}B_0x_2^2 \text{Vol}(S^2),\nn\\[2mm]
F_1 &= \partial_{x_3}(e^{4A-2\Phi})dx_1+\partial_{x_3}(e^{-2A} B_0)dx_2- \partial_{x_1}(e^{-4A})dx_3,\nn\\[2mm]
F_3 &=B\wedge F_1+x_2^2\bigg(\partial_{x_2}(e^{-4A})dx_3- \partial_{x_3}(e^{-2A}B_0)dx_1- \partial_{x_3}(e^{-8A+2\Phi}(1+ B_0^2))dx_2\bigg)\wedge \text{Vol}(S^2), \nn\\[2mm]
F_5&=0.
\end{align}
We note that $F_5=B\wedge F_3 = B\wedge  B\wedge F_1=0$ so there is no possibility of having any D3-brane charge, induced or otherwise - this suggests this case is an intersecting NS5-D5-D7-brane system, in line with our naive observations about the metric.
The Bianchi identities impose the following conditions 
\begin{equation}
\label{3.2_BI}
\begin{split}
&\partial_{x_3}^2(e^{4A-2\Phi} )+  \partial_{x_1}^2( e^{-4A})=0,\\[2mm]
&\partial_{x_3}^2(e^{-2A} B_0)+ \partial_{x_1}\partial_{x_2}(e^{-4A})=0,\\[2mm]
&\partial_{x_3}^2(e^{-8A+2\Phi}(1+B_0^2))+ \frac{1}{x_2^2}\partial_{x_2}(x_2^2\partial_{x_2}(e^{-4A}))=0.
\end{split}
\end{equation}
This is actually a restricted version of  the  PDEs that appear in \cite{Macpherson:2016xwk} appendix C, where one imposes that $\partial_{x_4}$ is an isometry and redefines the dilaton of that appendix as $e^{\Phi} \to e^{\Phi-A}$. This is because the system here is the T-dual of the intersecting NS5-D4-D6-brane system that appears in section 4.3 of  \cite{Macpherson:2016xwk}.\\

The equations \eqref{3.2_BPS} and \eqref{3.2_BI} are a highly nontrivial system of coupled PDEs in terms of 3 variables, which it seems very difficult to make progress with in general. For that reason we shall make some ans\"atze in the following subsections. Later on in section \ref{sec:ads6} we will also show how to find all AdS$_6$ solutions of IIB within this case, up to a cylindrical Laplace equation.

For what follows, it will be useful to express the metric and the fluxes using another set of coordinates. In this way it is also possible to simplify the large number of PDEs we have to solve.
The change of coordinates is given by:
\begin{equation}
y_1 = y_1(x_1,x_2,x_3) , \qquad \quad y_i = x_i  \quad i = 2,3,4
\end{equation}
where $y_1$ is defined by the following conditions:
\begin{equation}
\partial_{x_1} y_1 = e^{4A - 2 \Phi} , \qquad \partial_{x_2} y_1 = e^{-2A} B_0 , \qquad \partial_{x_3} y_1 = f(x_1,x_2,x_3) .
\end{equation}
The first two definitions are allowed thanks to \eqref{3.2_BPS1}, which is thus automatically solved. Consistency of the last definition with the other two implies the following conditions on $f$:
\begin{equation}
\partial_{x_1} f = \partial_{x_3} e^{4A - 2 \Phi} , \qquad \partial_{x_2} f = \partial_{x_3} e^{-2A} B_0 .
\end{equation}
Moreover we can fix the derivative of $f$ with respect to $x_3$ so that it satisfies the first two Bianchi identities in \eqref{3.2_BI}
\begin{equation}
\label{mink5_Dx3f}
\partial_{x_3} f = - \partial_{x_1} e^{-4A} + g(x_3)
\end{equation}
where $g$ is an arbitrary function that we will set to zero for simplicity. 
For clarity, we have the following change of coordinates:
\begin{equation}
\label{change:jacobian}
\begin{pmatrix}
\partial_{x_1} \\ \partial_{x_2} \\ \partial_{x_3}
\end{pmatrix} = 
\begin{bmatrix}
e^{4A-2\Phi} & 0 & 0 \\
e^{-2A} B_0 & 1 & 0 \\
f & 0 & 1
\end{bmatrix}
\begin{pmatrix}
\partial_{y_1} \\ \partial_{y_2} \\ \partial_{y_3}
\end{pmatrix}
\quad \text{,} \quad 
\begin{pmatrix}
\partial_{y_1} \\ \partial_{y_2} \\ \partial_{y_3}
\end{pmatrix} =
\begin{bmatrix}
e^{-4A+2\Phi} & 0 & 0 \\
-e^{-6A + 2 \Phi} B_0 & 1 & 0 \\
-e^{-4A+2\Phi}f & 0 & 1
\end{bmatrix}
\begin{pmatrix}
\partial_{x_1} \\ \partial_{x_2} \\ \partial_{x_3}
\end{pmatrix} 
\end{equation}
where $f$ is defined by
\begin{equation}
\label{3.2change_fcond}
\partial_{x_1} f = \partial_{x_3} e^{4A - 2 \Phi} , \qquad \partial_{x_2} f = \partial_{x_3} e^{-2A} B_0 , \qquad \partial_{x_3} f = - \partial_{x_1} e^{-4A} .
\end{equation}
This change of variables is just transforming the first element of the vielbein
\begin{equation}
v_1 = e^{3A - \Phi} d x_1 + B_0 e^{-3A + \Phi} d x_2 = e^{-A + \Phi} (d y_1 -  f d y_3 ) ,
\end{equation}
leaving the others unchanged, therefore the metrics reads:
\begin{equation}
\label{mink5_changedmetric}
d s^2_6 = e^{2A} d s^2 (S^1)+  e^{-6A + 2 \Phi} \bigg( d y_2^2+ y_2^2 d s^2 (S^2)  \bigg) + e^{-2A} d y_3^2+ e^{-2A + 2 \Phi} \big(d y_1 - f d y_3\big)^2 .
\end{equation}
Now we are left with just two PDEs to solve: first there is \eqref{3.2_BPS2}, which is somewhat simplified by our change of coordinates, becoming
\begin{equation}
\label{BPS2_change}
\frac{1}{y_2^2} \partial_{y_2} (y_2^2 e^{-6A + 2 \Phi} B_0) = \partial_{y_1} e^{-8A+2\Phi},
\end{equation}
second is the third Bianchi condition in \eqref{3.2_BI}, which we find it easier to compute by reimposing the Bianchi identity of $F_3$ in the new coordinate system. 
The fluxes become
\begin{align}
\label{mink5change_fluxes}
&B = e^{-6A + 2 \Phi} y_2^2 B_0 \text{Vol}(S^2) ,~~~F_1 = d f,~~~F_5 = 0, \\[2mm]
&F_3 = B \wedge F_1 + y_2^2  \text{Vol}(S^2) \wedge \Big[  \big( e^{-4A + 2 \Phi}f G + \partial_{ y_2} e^{-4A} + e^{-2A} B_0 \partial_{ y_1}e^{-4A} \big)d  y_3 - e^{-4A + 2 \Phi} G d  y_1,\nn \\[2mm]
&\quad - \big(\partial_{ y_3} e^{-8A+2\Phi} + f \partial_{ y_1} e^{-8A+2\Phi} + e^{-2A}B_0 \partial_{ y_3}(e^{-6A+2 \Phi}B_0) - e^{-2A}B_0 f \partial_{y_2}e^{-4A+2\Phi} \big)d y_2 \Big]\nn 
\end{align}
where
\begin{equation}
G = \partial_{y_3} (e^{-2A}B_0) + e^{-4A+2\Phi}f \partial_{y_2} e^{4A-2\Phi} + e^{-6A+2\Phi}B_0  \partial_{y_1} e^{4A-2\Phi} 
\end{equation}
and the Bianchi identity of $F_3$, that was \eqref{3.2_BI}, becomes
\begin{equation}
\label{BI3changed}
\begin{split}
&\frac{1}{y_2^2}\partial_{y_2} \big(y_2^2 ( e^{-4A + 2 \Phi}f G + \partial_{ y_2} e^{-4A} 
+ e^{-2A} B_0 \partial_{ y_1}e^{-4A}) \big) \\
&+ \partial_{y_3} \big( \partial_{ y_3} e^{-8A+2\Phi} + f \partial_{ y_1} e^{-8A+2\Phi} + e^{-2A}B_0 \partial_{ y_3}(e^{-6A+2 \Phi}B_0) - e^{-2A}B_0 f \partial_{y_2}e^{-4A+2\Phi} \big) = 0,
\end{split}
\end{equation}
in these new coordinates. This second condition however is still difficult to solve in full generality, so we shall now proceed to making some ans\"atze.

\subsubsection{Ansatz: $\partial_{y_1 y_3} x_1 = 0$}\label{sec:mink5anstz1}
The change of coordinates of the previous section allows us to express the functions $A,\Phi,B_0,f$ in terms of  derivatives of $x_1$ with respect to the new coordinates $\partial_{y_i}x_1$, as given in the first column of the inverse Jacobian in \eqref{change:jacobian}.
For example \eqref{BPS2_change} can be written as
\begin{equation}
\label{BPS2_change_x1}
\frac{1}{y_2^2} \partial_{y_2} (y_2^2 \partial_{y_2} x_1) + \partial_{y_1}(e^{-4A}  \partial_{y_1} x_1)=0 ,
\end{equation}
while the last relation of \eqref{3.2change_fcond} becomes:
\begin{equation}
\label{warpdef_change}
\partial_{y_1} e^{-4A} = \partial_{y_3}^2 x_1 - 2 \frac{\partial_{y_3} x_1}{\partial_{y_1}x_1} \partial_{y_1 y_3}^2 x_1 + \left(  \frac{\partial_{y_3}x_1}{\partial_{y_1}x_1} \right)^2 \partial_{y_1}^2 x_1 .
\end{equation}
These expressions can be simplified by imposing $\partial_{y_1 y_3} x_1 = 0$ which means that we can write
\begin{equation}
x_1 = F(y_1,y_2) + E(y_2 , y_3) .
\end{equation}
This ansatz allows us to integrate \eqref{warpdef_change} getting the following definition for the Minkowski warp factor
\begin{equation}
e^{-4A}=y_1 \partial_{y_3}^2 E - \frac{(\partial_{y_3} E)^2}{\partial_{y_1} F} + C(y_2,y_3),
\end{equation}
where $C$ is an arbitrary function. The dilaton is simply
\beq
e^{-2 \Phi} = e^{-4A}/\partial_{y_1} F.
\eeq
Using these assumptions \eqref{BPS2_change_x1} reads
\begin{equation}
\label{BPS2_change_1}
\frac{1}{y_2^2} \partial_{y_2} (y_2^2 \partial_{y_2}(F+E))+(y_1\partial_{y_3}^2 E+C )\partial_{y_1}^2 F+\partial_{y_1} F\partial_{y_3}^2 E=0 
\end{equation}
and the Bianchi identity \eqref{BI3changed} becomes
\begin{equation}
\frac{1}{y_2^2} \partial_{y_2} (y_2^2 \partial_{y_2}C)+(\partial_{y_1} F)^2 \partial_{y_1} \left(\frac{y_1}{\partial_{y_1} F}\right)\partial_{y_3}^2C=\partial_{y_3}^2(\partial_{y_3}E)^2+y_1^2\partial_{y_3}^4E\partial_{y_1}^2 F.
\end{equation}
Moreover, we get the following expression for the metric
\begin{equation}\label{eq:mink5anstz1metric}
d s^2 = e^{2A} d s^2 (\text{Mink}_5)+  e^{-2A}\partial_{y_1}F \big( d y_2^2+ y_2^2 d s^2 (S^2)  \big) + e^{-2A} d y_3^2+ e^{2A}\partial_{y_1}F \left(d y_1 + \frac{\partial_{y_3}E}{\partial_{y_1}F} d y_3\right)^2 ,
\end{equation}
while the fluxes read
\begin{equation}\label{eq:mink5anstz1flux}
\begin{split}
&B = -y_2^2 \partial_{y_2}(F+E) \text{Vol}(S^2) \, , \qquad F_1 = - d \left( \frac{ \partial_{y_3}E}{ \partial_{y_1}F} \right) \, , \qquad F_5 = 0 \\
&F_3 = y_2^2 \text{Vol}(S^2)\wedge \bigg[ \partial_{y_2}\left( \frac{\partial_{y_3}E}{ \partial_{y_1}F} \right) (\partial_{y_1}F d y_1 + \partial_{y_3}E d y_3) +\partial_{y_2} e^{-4A} dy_3 \\
&  \quad -  \left(\partial_{y_1}F \partial_{y_3}\bigg(\frac{e^{-4A}}{\partial_{y_1}F} - \frac{(\partial_{y_3}E)^2}{(\partial_{y_1}F)^2} \bigg) + \partial_{y_3}E\partial_{y_1} \bigg(\frac{e^{-4A}}{\partial_{y_1}F} + \frac{(\partial_{y_3}E)^2}{(\partial_{y_1}F)^2} \bigg) \right)  \partial_{y_1}F dy_2   \bigg]  \, .
\end{split}
\end{equation}

\subparagraph{Case 1}
\label{subcase1_mink5}
Now we can impose extra conditions to further simplify the expressions. The first thing we want to do is to set $\partial_{y_3}E = 0$ in order to eliminate the fibration in the metric. In this case we can absorb $E$ in $F$, i.e., set $E=0$ without loss of generality. 
The condition \eqref{BPS2_change_1} then becomes
\begin{equation}
\label{change_case1BPS}
\frac{1}{y_2^2} \partial_{y_2} (y_2^2 \partial_{y_2}F)+C\partial_{y_1}^2 F=0 
\end{equation}
and, taking the derivative with respect to $y_3$ of this we get:
\begin{equation}
\label{change_case1condition}
\partial_{y_3}C \partial_{y_1}^2 F = 0 \, .
\end{equation}
If we solve this constraint imposing $\partial_{y_3} C = 0$ then 
\begin{equation}
e^{-4A} = c_1 + \frac{c_2}{y_2} \, ,
\end{equation}
while $F$ has to solve the following PDE:
\begin{equation}\label{eq:masslesspde}
\frac{1}{y_2^2} \partial_{y_2} (y_2^2 \partial_{y_2} F) + e^{-4A} \partial_{y_1}^2 F = 0 .
\end{equation}
We can notice that this case is actually the T-dual of \cite{Macpherson:2016xwk} subsection 4.1.1 which is a system in massless IIA studied \cite{imamura-D8}.

However, we can solve \eqref{change_case1condition} also by imposing $\partial_{y_1}^2 F = 0$, and then, using \eqref{change_case1BPS}:
\begin{equation}\label{eq:case1F}
F = c_1 + \frac{c_2}{y_2} + y_1 \left(c_3 + \frac{c_4}{y_2}\right) 
\end{equation}
and we are left with only the Bianchi identity:
\begin{equation}\label{eq:D5NSbianchi}
\left(c_3 + \frac{c_4}{y_2}\right) \, \partial_{y_3}^2 e^{-4A} + \frac{1}{y_2^2} \partial_{y_2} (y_2^2 \partial_{y_2} e^{-4A}) = 0 .
\end{equation}
This brane system was studied in section 4.5 of \cite{youm},  and contains the solution of D5-brane ending on partially delocalised NS5-branes where $H_5 = e^{-4A}$ and $H_{NS} = \partial_{y_1}F = e^{-4A+2\Phi}$. 

We will now derive two more cases which generalize the system to include $F_1$ flux.
\subparagraph{Case 2} Another generalization of the first case can be obtained by the following ansatz:
\begin{equation}
\frac{1}{y_2^2} \partial_{y_2} (y_2^2 \partial_{y_2}E)=\partial_{y_3}^2 E=0.
\end{equation}
Using this condition, \eqref{BPS2_change_1} imposes
\begin{equation}
\frac{1}{y_2^2} \partial_{y_2} (y_2^2 \partial_{y_2}F)+C \partial_{y_1}^2 F=0, \qquad \partial_{y_3} C = 0
\end{equation}
while the Bianchi identity simply gives
\begin{equation}
\frac{1}{y_2^2} \partial_{y_2} (y_2^2 \partial_{y_2}C)=0.
\end{equation}
Summarizing, we get the following expressions:
\begin{equation}
\begin{split}
&E = \left( c_1 + \frac{c_2}{y_2}\right) + y_3 \left(c_3 + \frac{c_4}{y_2}\right) \, , \qquad  C = k_1 + \frac{k_2}{y_2}\\
&\frac{1}{y_2^2} \partial_{y_2} (y_2^2 \partial_{y_2}F)+C \partial_{y_1}^2 F=0,
\end{split}
\end{equation}
we recover the T-dual of \cite{imamura-D8} discussed in case 1 when $E=0$, the presence of the fibration in the metric \eqref{eq:mink5anstz1metric}, $ \left(d y_1 + \frac{\partial_{y_3}E}{\partial_{y_1}F} d y_3\right)^2$, acts as a source for a 7-brane as in section 6 of \cite{Bergshoeff:1996}. 
\subparagraph{Case 3}
\label{subcase2_mink5}
We can generalize what we saw in the previous case, leaving the fibration term which is associated with $F_1$ turned on. For example we can impose the following conditions:
\begin{equation}
\frac{1}{y_2^2} \partial_{y_2} (y_2^2 \partial_{y_2}F)=\partial_{y_1}^2 F=0
\end{equation}
in order to have
\begin{align}
&F = \left( c_1 + \frac{c_2}{y_2}\right) + y_1 \left(c_3 + \frac{c_4}{y_2}\right) \,, \\
&\frac{1}{y_2^2} \partial_{y_2} (y_2^2 \partial_{y_2}E)+\partial_{y_1} F\partial_{y_3}^2 E=0 \, , \label{Eeq_subansatz2}
\end{align}
while $C$ is determined by the following PDE coming from \eqref{BI3changed}
\begin{equation}
\label{Ceq_subansatz2}
\frac{1}{y_2^2} \partial_{y_2} (y_2^2 \partial_{y_2}C)+\partial_{y_1} F\partial_{y_3}^2 C= \partial_{y_3}^2 (\partial_{y_3}E)^2,
\end{equation}
we recover the D5-NS5 system of case 1 when $E=0$.

\subsubsection{Ansatz: $f = 0$ } 
Another way to get manageable solutions is to kill the fibrations in \eqref{mink5_changedmetric} by setting $f = 0$. However, in this case we also impose $g \neq 0$ in \eqref{mink5_Dx3f}, otherwise we will simply get
case 1 of \ref{subcase1_mink5}.
The metric now is diagonal
\begin{equation}
d s^2 = e^{2A} d s^2(\text{Mink}_5) + e^{-2A + 2 \Phi} d y_1^2 +  e^{-6A + 2 \Phi}  \bigg( d y_2^2+ y_2^2 d s^2 (S^2)  \bigg) + e^{-2A} d y_3^2 
\end{equation}
and the fluxes read:
\begin{subequations}
	\label{ansatz2_mink5}
	\begin{align}
	B_2 =& y_2^2 e^{-6A+2\Phi} B_0 \text{Vol}(S^2) \\
	F_1 =& - e^{4A - 2 \Phi} \partial_{y_1} e^{-4A} \, d y_3 \label{ansatz1_F1} \\
	F_3 =& B_2 \wedge F_1 +y_2^2  \text{Vol}(S^2) \wedge \left( \left( \partial_{y_2} e^{-4A} + e^{-2A} B_0\partial_{y_1} e^{-4A} \right) d y_3 - e^{-4A + 2 \Phi} \partial_{y_3} e^{-4A} \, d y_2   \right) \\
	F_5 =& 0 .
	\end{align}
\end{subequations}
The Bianchi identity becomes \eqref{BI3changed}:
\begin{equation}
\label{eq:ansatz1BI}
e^{-4A+2 \Phi} \partial_{y_3}^2 e^{-4A} + \frac{1}{y_2^2} \partial_{y_2} \left( y^2_2 \partial_{y_2} e^{-4A} \right) + \frac{1}{2} \partial_{y_1}^2 e^{-8A} =0 \, .
\end{equation}
Since, as said before, the $F_1 = 0$ case here is case 1 of \ref{subcase1_mink5}, we can start solving \eqref{eq:ansatz1BI} with the condition $\partial_{y_1}e^{-4A} \neq 0$, which implies also $g \neq 0$. 

We can start from \eqref{mink5_Dx3f}, which in the $y$ coordinates reads
\begin{equation}
\label{eq:ansatz1defPhi}
g e^{-4A + 2 \Phi} = \partial_{y_1} e^{-4A} .
\end{equation}
We can integrate this equation to get that the warping function can be written as
\begin{equation}
e^{-4A} = g F(y_1,y_2) + C(y_2,y_3)
\end{equation}
where $C$ is an arbitrary function while $F$ is defined by the condition $\partial_{y_1}F = e^{-4A + 2 \Phi}$.

Now we can use \eqref{BPS2_change}
\begin{equation}
\label{BPS2_changemod}
 \frac{1}{y_2^2} \partial_{y_2} \left( y^2_2 e^{-6A + 2 \Phi}B_0\right) = \frac{g}{2}\partial_{y_1}^2 F^2 + C \partial_{y_1}^2 F 
\end{equation}
to get some constraints on $F$ and $g$, indeed taking the derivative respect to $y_1$ and $y_3$ of this equation we get
\begin{equation}
\label{mink5splitcond}
g' \partial_{y_1} \left( \frac{\partial_{y_1}^2 F^2}{\partial_{y_1}^2 F} \right) = 0.
\end{equation}
Thus we have two possible solutions,  $g' = 0$ or $\partial_{y_1} \left( \partial_{y_1}^2 F^2/\partial_{y_1}^2 F \right) = 0$. The second case leads to a completely decoupled system of D5-D7-NS5-branes, so we will focus just on the first one.

Since the left-hand side of \eqref{BPS2_changemod} is $y_3$ independent, we have to impose that also the right-hand side is. We can obtain this in two ways: by setting $\partial_{y_3}C = 0$ or $\partial_{y_1}^2F = 0$.
\subparagraph{Case 1}
If $\partial_{y_3}C = 0$ this means that the warping function is an arbitrary function of $y_1,y_2$ which satisfy the following equation
\begin{equation}
\frac{1}{y_2^2} \partial_{y_2} \left( y_2^2 \partial_{y_2} e^{-4A} \right) + \frac{1}{2} \partial_{y_1}^2  e^{-8A} = 0. 
\end{equation}
This is the T-dual of the massive IIA case in section 4.1.2 of \cite{Macpherson:2016xwk}, which is a system studied also in \cite{imamura-D8}.

\subparagraph{Case 2}If $\partial_{y_1}^2F = 0$ we get the following expression for $A$:
\begin{equation}
e^{-4A} = g h(y_2) y_1 + C
\end{equation}
where $h$ is an arbitrary function. From the consistency relation
\begin{equation}
\partial_{y_1}(e^{-6A+2\Phi}B_0) = - \partial_{y_2}e^{-4A+2\Phi} = - h'
\end{equation}
we find that
\begin{equation}
e^{-6A+2\Phi}B_0 = - y_1 h'+b(y_2)
\end{equation}
and using this expression in \eqref{BPS2_changemod} we get
\begin{equation}
h = c_2 + \frac{c_1}{y_2} \quad \text{and} \quad e^{-6A+2\Phi}B_0 = \frac{1}{y_2^2} \left(c_1 y_1 +g c_1^2 y_2 + \frac{g}{3} c_2^2 y_2^3 +  g c_2 y_2^2 \right).
\end{equation}
Now we are left with just the Bianchi identity \eqref{eq:ansatz1BI} to solve
\begin{equation}
h \partial_{y_3}^2 C + \frac{1}{y_2^2} \partial_{y_2} \left( y^2_2 \partial_{y_2} C \right) + g^2 h^2 = 0;  
\end{equation}
if we perform the following transformation $C \rightarrow C - g^2 (c_1 c_2 y_2 + (c_2^2 y_2^2)/6 + c_1^2 \log y_2)$ we can absorb the non linear term in the last equation, that now simply reads
\begin{equation}
h \partial_{y_3}^2 C + \frac{1}{y_2^2} \partial_{y_2} \left( y^2_2 \partial_{y_2} C \right) = 0.
\end{equation}

\subsection{Case III: A new conformal Calabi-Yau system - Mink$_4$} \label{sec:conformalCalab}
For this case we solve \eqref{eq: physicalcondtion} with $\cos\zeta=0$ and additionally set $(\kappa_{\|}=1,~\kappa_{\perp}=0)$, meaning that the solutions in this class are of conformal Calabi-Yau type \cite{grana-polchinski,giddings-kachru-polchinski,Becker:1996gj,Dasgupta:1999ss}.  Generic values away from this choice share none of the physical characteristics of this class, which is no big surprise as $M_6$ would no longer be conformally Calabi-Yau.

Here it is possible to show that \eqref{eq:4dsusy1a}-\eqref{eq:4dsusy3b} are implied by 
\begin{subequations}
\begin{align}
d(e^{2A+2C-\Phi})+2 e^{2A+C-\Phi} v_2= d(e^{A-\frac{1}{2}\Phi}v_1)= d(e^{A}w)=0, \label{vielbein_kperp_z=0} \\[2mm]
B_2+B_0 v_1\wedge v_2=d(e^{2C}B_0)\wedge w\wedge\overline{w}=d(e^{-\Phi})\wedge w\wedge\overline{w}=0,  \label{NSNS_kperp_z=0}
\end{align}
\end{subequations}
where we have once more introduced potentials for the NS 3-form, $B_2$ and $B_0$,  as in \eqref{eq:LNSflux potential}.
As we did for the previous cases we solve \eqref{vielbein_kperp_z=0} by introducing local coordinates and using them to define the vielbein on $M_4$  as
\begin{equation}
v_1 =e^{-A + \frac{1}{2}\Phi} d x_1, ~~~ v_2 = - e^{-A +\frac{1}{2}\Phi} d x_2 , ~~~ w_1 = - e^{-A} d x_3, ~~~w_2 = - e^{-A} dx_4 ,~~~x_2=e^{A+C-\frac{1}{2}\Phi},
\end{equation}
without loss of generality. The conditions \eqref{NSNS_kperp_z=0} define part of the NS 2-form potential that lies orthogonal to $S^2$ and impose that
\beq\label{eq:functions}
e^{2C}B_0= g(x_3,~x_4),~~~~e^{-\Phi}= f(x_3,~x_4).
\eeq
The 10d metric takes the form
\begin{equation}\label{eq: conformalcalmet}
ds^2 =e^{2A}ds^2(\text{Mink}_4) + e^{- 2 A}  \bigg(\frac{1}{f}(d x_1^2 + d x_2^2 + x^2_2 d s^2(S^2)) + (d x_3^2 + d x_4^2) \bigg),
\end{equation}
we remark that $x_2$ is a radial coordinate such that the part of the metric spanned by $(x_1,~ x_2,~S^2)$ is warped $\mathbb{R}^4$  while the part spanned by $(x_3,~x_4)$ is warped $\mathbb{R}^2$ - so the Calabi-Yau metric $e^{2A}ds^2(M_6)$ can be viewed locally as a foliation of $T^4$ over $T^2$, although clearly the whole space does not generically exhibit such  isometries due to the coordinate dependence of $e^{2A}$. 

The 10 dimensional fluxes that follow from \eqref{RR_fluxes} are then given by
\begin{subequations}
\label{fluxesConformalCalab}
\begin{align}\small
B &= g {\cal C}_2,~~~ {\cal C}_2= \text{Vol}(S^2) + \frac{d x_1 \wedge d x_2}{x_2^2}, ~~~F_1 = \partial_{x_4} f d x_3 - \partial_{x_3} f d x_4, \\[2mm]
F_3 &=B\wedge F_1- \bigg(\partial_{x_4}(f g) dx_3-\partial_{x_3}(f g) dx_4\bigg)\wedge{\cal C}_2 ,\nonumber\\[2mm]
 F_5 &= B\wedge F_3- \frac{1}{2} B \wedge B \wedge F_1+x_2^2\bigg(\partial_{x_2}(e^{-4A})dx_1-\partial_{x_1}(e^{-4A})dx_2\bigg)\wedge dx_3\wedge dx_4\wedge \text{Vol}(S^2)\nn \\[2mm]
&+\frac{1}{2}\bigg(\partial_{x_4}(fg^2+ x_2^4f^{-1}e^{-4A} )dx_3-\partial_{x_3}(fg^2+ x_2^4f^{-1}e^{-4A} )dx_4\bigg)\wedge {\cal C}_2\wedge {\cal C}_2 + \text{Vol}_{4} \wedge d(e^{4A}f),
\end{align}
\end{subequations}
where $\text{Vol}_{4}$ is the volume form on Mink$_4$. All that is left to do is imposing the  Bianchi identities, from $F_1$ and $F_3$ we find the following PDEs
\begin{subequations}
\begin{align}
&\Box_2 f = 0 , ~~~  \Box_2 (fg) = 0,~~~ \Box_2= \partial_{x_3}^2+ \partial_{x_4}^2,\label{eq:CalabBianchi1}\\[2mm]
&\partial_{x_1}^2 (e^{-4A}) + \frac{1}{x_2^2} \partial_{x_2} ( x_2^2 \partial_{x_2} (e^{-4A}))  + \Box_2(e^{-4A} f^{-1}) +\frac{1}{x_2^4} \Box_2 (f g^2) = 0\label{eq:CalabBianchi2}.
\end{align}
\end{subequations}
This manifestly reduces to the PDEs governing section 4.2 of \cite{Macpherson:2016xwk} when we impose that $\partial_{x_4}$ is an isometry - 
this is because that section is a special case of this more general system, up to T-dual on $\partial_{x_4}$.

For the discussion of solutions to the Bianchi identities it is beneficial to rewrite the warp factor as
\begin{equation} \label{eq:conformalCalab_warp}
 \e^{-4A} = \frac{f}{x_2^2} \left[x_2^2 h - \left((\partial_{x_3} g)^2
+ (\partial_{x_4} g)^2\right) \right]
\end{equation}
with a function $h=h(x_1,x_2,x_3,x_4)$, so that \eqref{eq:CalabBianchi2}
turns into
\begin{equation}\label{eq:conformalCalabPDE_laplace}
f \left(\partial_{x_1}^2 h + \frac{1}{x_2^2} \partial_{x_2} ( x_2^2 \partial_{x_2} h) \right)  + \Box_2 h = \frac{1}{x_2^2} \Box_2  \left( (\partial_{x_3} g)^2 + (\partial_{x_4} g)^2 \right),
\end{equation}
which makes the structure of a Laplace equation more evident.  The generic case remains quite complicated, so we consider some special cases in the following  which yield more explicit insights.

\subsubsection{Ansatz $F_3=H=0$ (constant $g$)}
The right-hand side of \eqref{eq:conformalCalabPDE_laplace} vanishes 
for a constant function $g$ which can be set to zero without loss of 
generality. The only non-vanishing fluxes are $F_1$ and $F_5$ and with
$\e^{-4A}=f h$ from \eqref{eq:conformalCalab_warp} 
the expressions turn into
\begin{subequations}
\begin{align}
&d s^2_{10} = \frac{1}{\sqrt{fh}} d^2s (\text{Mink}_4) +
  \frac{\sqrt{h}}{\sqrt{f}}(d x_1^2 + d x_2^2 + x^2_2 d s^2(S^2)) + \sqrt{f h} (d x_3^2 + d x_4^2), \quad e^{-\Phi} = f, 
\label{eq:conformalCalab_constant_g_metric}\\[2mm]
 &F_1 = \partial_{x_4} f d x_3 - \partial_{x_3} f d x_4, \\[2mm]
 &F_5 =\text{Vol}_{4} \wedge dh^{-1} + x_2^2  \text{Vol}(S^2) \wedge \big( \eps_{ij}\partial_{x_i}h \, dx_j \big) \wedge \big( f dx_3\wedge dx_4 +  d x_1 \wedge d x_2 \big), \\[2mm]
\label{eq:conformalCalab_constant_g_bianchi}
&f \left[\partial_{x_1}^2 h +  \frac{1}{x_2^2} \partial_{x_2} ( x_2^2 \partial_{x_2} h)\right]  + \Box_2 h  = 0,  \qquad   \Box_2 f = 0.
\end{align}
\end{subequations}
The resulting 10 dimensional metric \eqref{eq:conformalCalab_constant_g_metric}
resembles the setup of intersecting D3-D7-branes, see for instance Eqs. (13)
and (14) in \cite{Behrndt:1996pm} or the T-dual of Eqs. (7.3)-(7.6) in \cite{imamura-D8}. In particular, we have re-derived the D3-D7 case of the system used to study localised Dp-branes in the world volume of D(p+4)-branes - see section 4.3 in \cite{youm}, but note that the D3-D7 system is exactly where the techniques of this paper fail to find a solution.

\subsubsection{Ansatz $F_1=0$ (constant dilaton) and special choices 
for $g$}
Another simplification of  the setup
occurs for a constant dilaton (without loss of generality $f=1$)
and special choices of $g$ such that the right-hand side of
 \eqref{eq:conformalCalabPDE_laplace} simplifies; it vanishes
for the function $g= c_1 x_3 + c_2 x_4$. The resulting scenario is
then described by  
\begin{subequations}\label{eq:simplesystem}
\begin{align}
&d s^2_6 = e^{-2A} (d x_1^2 + d x_2^2 + x^2_2 d s^2(S^2) + d x_3^2 + d x_4^2), \quad \Phi = 0, \quad e^{-4A} = h - \frac{c_1^2 + c_2^2}{x_2^2}, \\[2mm]
&B = (c_1 x_3 + c_2 x_4) {\cal C}_2,  \quad F_3 =  - \big(c_2 dx_3- c_1 dx_4\big)\wedge{\cal C}_2 \quad  {\cal C}_2= \text{Vol}(S^2) + \frac{d x_1 \wedge d x_2}{x_2^2}, \\[2mm]
&F_5 = \text{Vol}_{4} \wedge d(e^{4A})- \frac{x_2^2}{6} \eps_{ijkl} \partial_{x_i} \big( e^{-4A} \big) d x_j \wedge d x_k \wedge d x_l \wedge \text{Vol}(S^2), \\[2mm]
& \label{eq:conformalCalab_const_dil_bianchi} 
\partial_{x_1}^2 h + \frac{1}{x_2^2} \partial_{x_2} ( x_2^2 \partial_{x_2} h)  + \Box_2h  = 0.
\end{align}
\end{subequations}
The choice $g= c_1 x_3^2 - c_1 x_4^2 +c_2 x_3 x_4$ yields similar expressions. More precisely, compensating the resulting RHS of \eqref{eq:conformalCalabPDE_laplace}
by including $c^2\mathrm{log}(x_2)$ 
with $c^2= 4c_1^2 + c_2^2 $ into $h$, 
one obtains
\begin{subequations}
\begin{align}
&d s^2_6 = e^{-2A} (d x_1^2 + d x_2^2 + x^2_2 d s^2(S^2) + d x_3^2 + d x_4^2), 
\quad e^{-4A} = h - c^2\frac{x_3^2 + x_4^2}{x_2^2} + 4 c^2 \log x_2 \\[2mm]
&B = (c_1 x_3^2 - c_1 x_4^2 +c_2 x_3 x_4) {\cal C}_2,
  \quad F_3 =  -  \big((c_2 x_3 - 2c_1 x_4) dx_3- (c_2 x_4 - 2c_1 x_3)  dx_4\big)\wedge{\cal C}_2, 
\end{align}
\end{subequations}
with $\Phi=0$ and $\mathcal{C}_2$ defined as before and exactly
the same expressions for the Bianchi identity and the flux $F_5$
as in the previous case. Thus, 
one has to solve  \eqref{eq:conformalCalab_const_dil_bianchi}
which is the Laplace equation on the internal space $M_6$, subject to
the $S^2$ isometry by our fundamental ansatz. 
 
In both cases the metric looks formally like the back reaction of a D3 with harmonic function $e^{-4A}$, however we have for $c_1,c_2 \neq 0$ that $e^{-4A}$ is not a harmonic function but the difference between a harmonic function $h$ and a not-harmonic term - something similar (albeit less general) was found in section 4.2.1 of \cite{Macpherson:2016xwk}.
One can derive several explicit solutions for $h$ by imposing  spherical symmetry with respect to the factor described by $(x_1,x_3,x_4)$.
Plugging these solutions into the expressions for the warp factor $\e^{-4A}$, some restrictions of the domain will arise as this must remain
positive. It is instructive to study the ``boundaries'' for some solutions $h$ in more detail. This procedure is illuminated in 
section~\ref{sec:examples_compact}, where we discuss a compact example.

\subsection{Generalized systems}
In this section we allow the functions $\kappa_{\|},~\kappa_{\perp}$  to take generic values away from the specific choices of sections  \ref{sec:D5}-\ref{sec:conformalCalab}. We find two distinct branches of solution that generalize cases I and II of sections  \ref{sec:D5} and \ref{sec:mink5system} to systems with more complicated geometry and additional flux, but governed by the same PDEs. These branches are distinguished by whether \eqref{eq: physicalcondtion} is solved with $\cos\zeta=0$ or $H_1=0$. Case III of section \ref{sec:conformalCalab} has no such generalization which we believe is because it lacks the additional $U(1)$ isometries the others enjoy. Specifically we anticipate that the two branches of solution we find can be generated from case I and case II under the combined actions of formal U-duality \cite{Maldacena:2009mw,Caceres:2011zn} and T-s-T like \cite{Lunin:2005jy} transformations that require the additional $U(1)$ isometries to work - similar behaviour was observed in \cite{Macpherson:2016xwk}.  However we delay confirmation of this hypothesis until \cite{toappearI} and for now satisfy ourselves with completing our classification by presenting the distinct branches.
\subsubsection{Case I Generalization}\label{sec: Gen I}
For our penultimate case we choose to solve \eqref{eq: physicalcondtion} with $H_1=0$ and assume also that $\kappa_{\|}\neq 0$ and  $\cos\zeta\neq 0$, as equality was dealt with in section \ref{sec:mink5system}.
After some work simplifying the various expressions we find that  \eqref{eq:4dsusy1a}-\eqref{eq:4dsusy3b} are implied by the necessary and sufficient supersymmetry conditions
\begin{subequations}
\begin{align}
&B_0=d(e^{2A-\Phi}\kappa_{\|}\cos\zeta )= d(e^{2A-\Phi}\kappa_{\perp})= d(\kappa_{\|}e^{-\Phi}\sin\zeta)=0,\label{B0zerogenBPSa}\\[2mm]
&d(e^{2A+2C})+ 2 e^{2A+C}v_2= d(e^{A-\Phi}k_2)= d\left(\frac{e^{A}}{\kappa_{\|}}(\text{Re}k_1- \text{Im}k_1 \tan\zeta)\right)=0,\label{B0zerogenBPSb}\\[2mm]
& d\left(\frac{e^{3A}}{\kappa_{\|}}\right)\wedge v_2\wedge(\text{Re}k_1- \text{Im}k_1 \tan\zeta)=0\label{B0zerogenBPSc},\\[2mm]
&\kappa_{\|}B_2= -\tan\zeta \kappa_{\|}(\text{Im}k_1\wedge \text{Re} k_1+\text{Im}k_2\wedge \text{Re} k_2)- \tan\zeta \kappa_{\perp}(\text{Im}k_2\wedge\text{Re}k_1- \text{Im}k_1\wedge \text{Re}k_2)\nn\\[2mm]
&+\kappa_{\perp}(\text{Im}k_1\wedge \text{Im}k_2-\text{Re}k_1\wedge \text{Re}k_2)\label{B0zerogenBPSd}
\end{align}
\end{subequations}
where we have defined
\beq\label{eq:kdef}
k_1= \kappa_{\|}v_1- \kappa_{\perp} w,~~~ k_2=\kappa_{\perp}v_1+ \kappa_{\|} w,
\eeq
to ease presentation and we remind the reader that $B_2$ is a potential for the part of the NS 3-form that lies orthogonal to $S^2$ satisfying $dB_2 = H_3$.
The conditions in \eqref{B0zerogenBPSa} give $\kappa_{\|}$, $\kappa_{\perp}$, $\zeta$ and $e^{A}$ as functions of $e^{-2\Phi}$
\beq\label{eq:secondlastbps1solve}
e^{2A}= \sqrt{\frac{c_1^2+ c_2^2}{1-c_3 e^{2\Phi}}
}e^{\Phi},~~~\cos\zeta= c_1\sqrt{\frac{1-c_3 e^{2\Phi}}{c_1^2+ c_2^2 c_3 e^{2\Phi}}},~~~\kappa_{\perp}= c_2\sqrt{\frac{1-c_3e^{2\Phi}}{c_1^2+c_2^2}},
\eeq
where $c_i$ are integration constants - notice that when $c_3=0$, which is when $\zeta=0$,  $\kappa_{\perp}$ becomes constant.

The conditions \eqref{B0zerogenBPSb} define the vielbein via
\beq\label{eq:secondlastveil}
v_2=- e^{-A}dx_2,~~~ k_2= e^{-A+\Phi}(dx_3+ i dx_4),~~ \text{Re}k_1= e^{-A}\left(\kappa_{\|}dx_1- \frac{\kappa_{\perp}}{\kappa_{\|}}e^{-A+\Phi}\tan\zeta dx_4\right),~~~x_2= e^{A+C}\nn,
\eeq
while  \eqref{B0zerogenBPSc} imposes that $\frac{e^{3A}}{\kappa_{\|}}$ depends on $x_1,x_2$ only which, given \eqref{B0zerogenBPSa}, makes $\partial_{x_3}$ and $\partial_{x_4}$ isometries of the metric. Contrary to case I however, these isometry directions do not share a common warp factor with Mink$_4$ and the $\partial_{x_4}$ direction is fibred over some base - thus we do not have an enhancement to Mink$_6$ generically. The 10D metric can be written as
\begin{subequations}\label{eq: caseIIgenmetric}
\begin{align}
ds^2 &=e^{2A}ds^2(\text{Mink}_4)+ e^{-2A+2\Phi}ds^2(\tilde{T}^2)+ e^{-2A}\bigg(\frac{\kappa_{\|}^2}{1+ \kappa_{\perp}^2\tan^2\zeta}dx_1^2+dx_2^2+ x_2^2 ds^2(S^2)\bigg),\label{eq: caseIIgenmetrica}\\[2mm]
ds^2(\tilde{T}^2)&= dx_3^2+ \frac{1+ \kappa_{\perp}^2\tan^2\zeta}{\kappa_{\|}^2}\bigg(dx_4- \frac{e^{-\Phi}\kappa_{\|}^2 \kappa_{\perp}\tan\zeta}{1+\kappa^2_{\perp}\tan^2\zeta}dx_1\bigg)^2,\label{eq: caseIIgenmetricb}
\end{align}
\end{subequations}
however, one should bare in mind that this is less complicated than it at first sight appears because
\beq
\frac{\kappa_{\|}^2}{1+ \tan^2\zeta \kappa_{\|}^2}= \frac{c_1^2}{c_1^2+ c_2^2}
\eeq
due to \eqref{eq:secondlastbps1solve}. This means that the final term in \eqref{eq: caseIIgenmetrica} is warped $\mathbb{R}^4$ up to rescaling $x_1$, and so the internal metric is locally a $T^2$ bundle over this warped $\mathbb{R}^4$.

We follow the prescription of section \ref{sec:D5} to establish what 10 dimensional fluxes these solutions support, starting from \eqref{eq:4dfluxa}-\eqref{eq:4dfluxb} and then using the vielbein of \eqref{eq:secondlastveil} to take the Hodge dual. After significant massaging they can be expressed as
\begin{subequations}\label{eq:caseIgenflux}
\begin{align}
B&= -e^{-2A+\Phi}\kappa_{\perp} dx_1\wedge dx_3 +e^{-2A+2\Phi}\tan\zeta dx_3\wedge dx_4,~~~F_1 = 0, \\[2mm]
F_3 &=  \frac{e^{-2A+\Phi}}{\cos\zeta} x_2^2\bigg(\partial_{x_2}(\kappa_{\|}^2e^{-2\Phi})dx_1- \partial_{x_1}(e^{-2\Phi})dx_2\bigg)\wedge \text{Vol} (S^2) ,\\[2mm]
F_5 &= \kappa_{\|}\sin\zeta e^{-\Phi}\text{Vol}_4\wedge d( e^{4A})+ B_2 \wedge F_3\\[2mm]
+&x_2^2d\left(\frac{\kappa_{\perp}}{\kappa_{\|}\cos^3\zeta}e^{-4A}\right)\wedge dx_2\wedge dx_3\wedge \text{Vol}(S^2).
\end{align}
\end{subequations}
The striking thing about these fluxes is that only $F_3$ does not manifestly give rise to its Bianchi identity when we act with the exterior derivative - given that $F_1=0$, $F_3$ should be closed, ensuring that it  leads to the PDE
\beq
\partial_{x_1}^2(e^{-2\Phi})+ \frac{c_1^2}{c_1^2+ c_2^2}\frac{1}{x_2^2}\partial_{x_2}(x_2^2\partial_{x_2}(e^{-2\Phi}))=0
\eeq
which is the same as the D5-brane Laplacian of \eqref{eq:D5Laplacian}, up to rescaling $x_1$, if we once more define $e^{-2\Phi}= h(x_1,x_2)$. Thus any $SO(3)$ preserving solution of the D5-brane Laplacian is also a solution of this more general system. This suggests that some form of duality is at play. As partial evidence for this consider the following: If we set $\kappa_{\perp}=0$ in  \eqref{eq:secondlastbps1solve}, \eqref{eq: caseIIgenmetric} and  \eqref{eq:caseIgenflux} we see that $T^2$ is no longer fibred and the metric and fluxes are precisely what one would expect to find after performing U-duality (of the kind discussed in \cite{Caceres:2011zn}) on a Mink$_4$ factor of the D5-brane. What duality is at play in full generality remains to be seen, however a sequence of U-duality and T-s-T transformations seems quite likely.
\subsubsection{Case II Generalization}\label{sec: Gen II}
For this final case we solve \eqref{eq: physicalcondtion} by fixing $\cos\zeta=0$ and additionally assume that  $\kappa_{\perp}\neq 0$, as we have already dealt with the contrary in section \ref{sec:D5}. We find a class of solutions that reduce to those in section \ref{sec:mink5system} when $\kappa_{\perp}=1$, and are in general governed by the same PDEs as that section, but with modified metric and fluxes.
The necessary and sufficient conditions for supersymmetry can be succinctly written as
\begin{subequations}
\begin{align}
& B_2=  \frac{\kappa_{\|}}{\kappa_{\perp}}\text{Re}k_1\wedge \text{Re}k_2,~~~d\left(e^{-2A+\Phi/2} \frac{\sqrt{\kappa_{\|}}}{\kappa_{\perp}}\right)=d(e^{3A+C-\Phi}\kappa_{\perp})+ e^{3A-\Phi}\kappa_{\perp}v_2=0,\label{eq:BPSlasta}\\[2mm]
& d\left(\frac{e^{A}}{\kappa_{\perp}}\text{Re}k_1\right)=d\left(\frac{e^{-A}}{\kappa_{\perp}}\text{Im}k_1\right)=d\left(\frac{e^{-3A+\Phi}}{\kappa_{\perp}^2}(\text{Re}k_2+ B_0 \kappa_{\perp} v_2)\right)=0,\label{eq:BPSlastb}\\[2mm]
&d(e^{A-\Phi}\text{Re}k_2)\wedge \text{Im}k_2=d(e^{A+2C-\Phi}(B_0\text{Re}k_2-\kappa_{\perp}v_2))\wedge \text{Im}k_2=0\label{eq:BPSlastc}.
\end{align}
\end{subequations}
where we once again introduce $k_i$ as defined in \eqref{eq:kdef} to ease presentation and introduced potentials $B_0,~B_2$ for the components of the NS 3-form sitting parallel and orthogonal to $S^2$ respectively as in \eqref{eq:LNSflux potential}.
We solve most of \eqref{eq:BPSlasta}-\eqref{eq:BPSlastb} without loss of generality by defining the various one-forms that appear in terms of local coordinates as
\begin{align}
&v_2=- \frac{e^{-3A+\Phi}}{\kappa_{\perp}}dx_2,~~~x_2= e^{3A+C-\Phi}\kappa_{\perp},\\[2mm]
&\text{Re}k_1= -e^{A}\kappa_{\perp}dx_4,~~~\text{Im}k_1=- e^{-A}\kappa_{\perp}dx_3,~~~ \text{Re}k_2=e^{3A-\Phi}(\kappa_{\perp}^2dx_1+ B_0 e^{-6A+2\Phi} dx_2),\nn
\end{align}
which in turn define the vielbein on $M_4$ through \eqref{eq:kdef}. What remains defines the potential $B_2$ and imposes that the functions $\kappa_{\perp}$ and $\kappa_{\|}$ obey the constraint
\beq
\kappa_{\|}= c e^{4A-\Phi} \kappa_{\perp}^2,
\eeq
for $c$ an arbitrary constant, which fixes the form of $\kappa_{\|}$ and $\kappa_{\perp}$ given that they must also solve $\kappa_{\|}^2+\kappa_{\perp}^2=1$ . We are left with only \eqref{eq:BPSlastc}, which implies PDEs that are similar to those \eqref{3.2_BPS} of case II
\beq
\partial_{x_2} (e^{4A - 2 \Phi}\kappa_{\perp}^2) = \partial_{x_1} (e^{-2A} B_0) ,~~~\frac{1}{x^2_2} \partial_{x_2} (x_2^2 e^{-2A} B_0) = \partial_{x_1} \left(\frac{e^{-8A+2 \Phi}}{\kappa_{\perp}^4}(1+\kappa_{\perp}^2 B_0^2 ) \right),\label{eq:BPSPDElast}
\eeq
with  $\partial_{x_4}$ one more an isometry direction. Unlike case II the isometry direction no longer has the same warping as Mink$_4$, so there can be no enhancement to Mink$_5$ and so it makes more sense for it to be defining an $S^1$ locally.

The 10D metric then takes the form  
\begin{align}
ds^2&= e^{2A}ds^2(\text{Mink}_4)+e^{2A}\kappa_{\perp}^2ds^2(S^1)+ \frac{e^{-6A+2\Phi}}{\kappa_{\perp}^2}\bigg(dx_2^2+ x_2^2 ds^2(S^2)\bigg)\nn\\[2mm]
&+e^{-2A}dx_3^2+e^{6A-2\Phi}\kappa_{\perp}^4\bigg(dx_1+ \frac{e^{-6A+2\Phi}B_0}{\kappa_{\perp}^2}dx_2\bigg)^2\nn,
\end{align}
with $ds^2(S^1)=dx_4^2$ - note that in the limit $\kappa_{\perp} \to 1$ this precisely reproduces the metric of case II \eqref{eq: case II metric} . 

For the final time we follow the procedure of section \ref{sec:D5} to extract the 10D fluxes from \eqref{eq:4dfluxa}-\eqref{eq:4dfluxb} which, after judicious use of \eqref{eq:BPSPDElast}, can be put in the form
\begin{align}
B&= \frac{1}{\kappa_{\perp}^2}e^{-6A+ 2\Phi}x_2^2B_0\text{Vol}(S^2) + e^{4A-\Phi}\kappa_{\|}(\kappa_{\perp}^2 dx_1 + e^{-2A+\Phi}B_0 dx_2)\wedge dx_4,\\[2mm]
F_1 &= \partial_{x_3}(e^{4A-2\Phi}\kappa_{\perp}^2)dx_1+\partial_{x_3}(e^{-2A} B_0)dx_2- \partial_{x_1}\left(\frac{e^{-4A}}{\kappa_{\perp}^2}\right)dx_3,\nn\\[2mm]
F_3 &=B\wedge F_1+x_2^2\bigg[\partial_{x_2}\left(\frac{e^{-4A}}{\kappa_{\perp}^2}\right)dx_3- \partial_{x_3}(e^{-2A}B_0)dx_1- \partial_{x_3}\left(\frac{e^{-8A+2\Phi}}{\kappa_{\perp}^4}(1+ \kappa_{\perp}^2 B_0^2)\right)dx_2\bigg]\wedge \text{Vol}(S^2)\nn\\[2mm]
&- d\big(\kappa_{\|}e^{-\Phi}\big)\wedge dx_3\wedge dx_4 \nn\\[2mm]
F_5&=\text{Vol}_4\wedge d(e^{4A-\Phi} \kappa_{\|})+B\wedge F_3-\frac{1}{2} B\wedge B\wedge F_1+d\left(x_2^2 \frac{\kappa_{\|}}{\kappa_{\perp}^2} e^{-6A+\Phi}B_0\right)\wedge dx_3\wedge dx_4\wedge \text{Vol}(S^2)\nn.
\end{align}
The parts of these expressions that do not manifestly give rise to the Bianchi identites of the RR fluxes when we act with the exterior derivative then impose
\begin{align}\label{eq:bianchipdelast}
&\partial_{x_3}^2(e^{4A-2\Phi} \kappa_{\perp}^2)+  \partial_{x_1}^2\left( \frac{e^{-4A}}{\kappa_{\perp}^2}\right)=0,~~~\partial_{x_3}^2(e^{-2A} B_0)+ \partial_{x_1}\partial_{x_2}\left(\frac{e^{-4A}}{\kappa_{\perp}^2}\right)=0,\nn\\[2mm]
&\partial_{x_3}^2\left(\frac{e^{-8A+2\Phi}}{\kappa_{\perp}^4}(1+\kappa_{\perp}^2B_0^2)\right)+ \frac{1}{x_2^2}\partial_{x_2}\left(x_2^2\partial_{x_2}\left(\frac{e^{-4A}}{\kappa_{\perp}^2}\right)\right)=0,
\end{align}
which reduces to the Bianchi identity PDEs of case II \eqref{3.2_BI} when $\kappa_{\perp}=1$. Actually we can make a much stronger statement than this: if we redefine the physical field as
\beq
e^{2A}= \frac{1}{\kappa_{\perp}}e^{2A^{II}},~~~ B_0= \frac{1}{\kappa_{\perp}}B_0^{II},~~~ e^{\Phi}= e^{\Phi^{II}},
\eeq
we can notice that this makes all dependence on $\kappa_{\perp}$ in \eqref{eq:BPSPDElast} and \eqref{eq:bianchipdelast} mutually cancel, leaving us with the equivalent expressions in section \ref{sec:mink5system}.
It seems likely then that solutions in this class can be generated by some sort of duality acting on the class of Mink$_5$ solutions of section \ref{sec:mink5system} - this example of duality seems a little more mysterious than that of the previous section - On the one hand it does not appear to be simply a U-duality while on the other hand it does not contain two $U(1)$ isometries on which to perform a standard T-s-T transformation. We do, however, know that the case II solutions descend via T-duality from a system in IIA \cite{Macpherson:2016xwk}, after imposing a $U(1)$ isometry on that solutions internal space - it is perhaps possible to define this $U(1)$ differently before dualising and arrive at the system we present here.\\

Let us now look at some specific examples which follow from our classification.
\section{Examples:}\label{sec:examples}
In this section, we consider some solution types in more detail, starting with an example of a conformal Calabi-Yau compactification in section \ref{sec:examples_compact}. We then consider AdS and linear dilaton  solutions in sections \ref{sec:AdSexamples} and \ref{sec:linearD}, respectively. 
\subsection{Compact solution}\label{sec:examples_compact}
We would like to show that one can find compact Minkowski solutions within our classification. This necessitates the inclusion of O-planes \cite{maldacena-nunez}. The simplest way to construct a solution consistent with the simultaneous requirements of small curvature and orientifold charge is from the conformal Calabi-Yau case \ref{sec:conformalCalab}, in particular from  \eqref{eq:simplesystem} where the Minkowski warp factor is the difference of a harmonic function and a non harmonic term - this solution is a generalization of one found in \cite{Macpherson:2016xwk}. We apply the ansatz
\beq
h= 4\frac{(a-x_2)}{x_2},~~~c_1= a\sin\phi_0\sin\theta_0,~~~c_2= a\cos\phi_0\sin\theta_0,~~~f=1,~~~g= c_1x_3+c_2 x_4,
\eeq
where $0<\theta_0<\frac{\pi}{2}$ so that the warp factor becomes
\beq
e^{-4A}= 4\frac{(a_+-x_2)(x_2-a_-)}{x_2^2},~~~a_{\pm}= \frac{a}{2}(1\pm \cos\theta_0).
\eeq
The need for this to stay positive bounds $a_-<x_2<a_+$ - this is also the reason to parametrise $c_1,c_2$ as we have - and the metric and fluxes and dilaton are
\begin{align}
ds^2&= e^{2A}ds^2(\text{Mink}_4) + e^{-2A}\bigg(dx_1^2+dx_2^2+dx_3^2+dx_4^2+x_2^2 ds^2(S^2)\bigg),~~~e^{\Phi}=1\nn\\[2mm]
F_1&=0,~~~H = \big(c_1 dx_3+c_2 dx_4\big)\wedge \big(\frac{dx_1\wedge dx_2}{x_2^2}+ \text{Vol}(S^2)\big),\nn\\[2mm]
F_3&=\big(c_1 dx_4-c_2 dx_3\big)\wedge \big(\frac{dx_1\wedge dx_2}{x_2^2}+ \text{Vol}(S^2)\big),\nn\\[2mm]
F_5&= -\frac{2ax_2(a\sin^2\theta_0-2x_2)}{(a^2 \sin^2\theta_0+4x_2^2-4a x_2)^2}\text{Vol}_4\wedge dx_2+ \frac{2a(a\sin^2\theta_0-2x_2)}{x_2}dx_1\wedge dx_3 \wedge dx_4 \wedge \text{Vol}(S^2),
\end{align}
where $\partial_{x_1},\partial_{x_3},\partial_{x_4}$ are isometry directions that we can take to be describing periodic directions.
The metric is regular for $x_2>0$ and between  $x_2=a_{\pm}$ - at the extrema it is singular tending to
\beq
ds^2= \frac{a_{\pm}}{\sqrt{a_+-a_-}\sqrt{\pm a_{\pm}\mp x_2}}ds^2(\text{Mink}_4)+\frac{\sqrt{a_+-a_-}\sqrt{\pm a_{\pm}\mp x_2}}{a_{\pm}}\bigg(dx_1^2+dx_2^2+dx_3^2+dx_4^2+ a_{\pm}^2 ds^2(S^2)\bigg)\nn
\eeq
which we recognise as the behaviour of an O3-plane smeared  on the $T^3$ spanned by $(x_1,x_3,x_4)$. Smeared O-planes are not physically valid,  we can however  cure this issue by T-dualising on $T^3$ which maps us to a system of localised O6-planes which can be interpreted as a Mink$_4$ vacuum with a compact internal space that is  $T^3$ fibred over $S^2$ times a finite interval.

\subsection{Solutions with AdS factors}\label{sec:AdSexamples}
In this section we would like to find all the AdS solutions compatible with our classification.
Any supergravity solution with an AdS$_{d+1}$ factor admits a realisation containing a Mink$_{d}$ factor, this parametrisation is called the Poincar\'e patch. As such the classification of warped AdS$_{d+1}$ solutions in  type IIB  is contained within the classification of warped Mink$_{d}$ solutions, one simply needs to make an ansatz such that
\beq
e^{2A_d}ds^2(\text{Mink}_d)+ ds^2(\text{M}_{10-d})= e^{2A_{d+1}} ds^2(\text{AdS}_{d+1})+ ds^2(\text{M}_{9-d}),
\eeq
and similarly for the fluxes. At the level of the metric it is clear that this requires taking $e^{2A_d}= r^2 e^{2A_{d+1}}$ where $r$ is the AdS radius, and arranging for the internal metric to decompose as $ds^2(\text{M}_{10-d})=  e^{2A_{d+1}}\frac{dr^2}{r^2} + ds^2(\text{M}_{9-d})$ so that we realise $AdS_{d+1}$ as the Poincar\'e patch. However when one generically infers an AdS classification from Minkowski one has limited prior knowledge about the specific form of the internal metric, so it is often better to do this second step at the level of spinors (see for example 	\cite{Apruzzi:2013yva,Apruzzi:2015zna}) - but this is much easier to do at the level of geometry if one has a local description.\\

In section \ref{sec: classification} we classified all\footnote{Up to the assumption of equal spinor norm, which is actually a requirement for the AdS solutions we consider.} Mink$_4\times S^2$ solutions of type IIB and were able to give local expressions in all cases - those that are potentially compatible with $AdS_{7,6,5}$ are sections \ref{sec:D5}, \ref{sec:mink5system}, \ref{sec:conformalCalab} respectively as these contain the correct Minkowski factors. Naively one might assume that the more general cases of section \ref{sec: classification}  would be compatible with AdS$_5$ as they contain Mink$_4$  factors, however the additional $U(1)$ isometries preclude this\footnote{Specifically the additional isometry direction are necessarily  warped by the - would be - AdS radius. This is at odds with the requirement that the geometry should be a product of $AdS$ and some distinct internal metric - so the $SO(2,3)$ isometry we are attempting to impose is broken.}.  We can actually disregard the possibility of AdS$_7$ solutions in a similar way: \ref{sec:D5} contains only the D5-brane, which can clearly never globally preserve the isometry group of AdS$_7$ due to the dilaton dependence on the D5 warp factor. This restricts our focus to all\footnote{All AdS$_6$ solutions are exactly $\mathcal{N}=1$ (by which we mean half-maximal supersymmetry in $AdS_6$ with 16 real supercharges) when they preserve supersymmetry and all realise their $SU(2)$ R-symmetry with a round $S^2$ at least locally \cite{Apruzzi:2014qva}.}  supersymmetric AdS$_6$ solutions of type IIB  which we cover in section \ref{sec:ads6};  and all\footnote{By $\mathcal{N}=2$ AdS$_5$ solutions we mean half-maximally supersymmetric solutions, which should have an $SU(2)\times U(1)$ R-symmetry that should be realised geometrically - as far as we are aware though and unlike the AdS$_6$, there is no proof that the $SU(2)$ factor needs to come from a round $S^2$} $\mathcal{N}=2$ supersymmetric  AdS$_5\times S^2$ solutions of type IIB  which will be dealt with in section \ref{sec:ads5}.

\subsubsection{AdS$_6$}\label{sec:ads6}
In this section we will recover all supersymmetric AdS$_6$ solutions  of IIB. These were already classified in \cite{Apruzzi:2014qva} and confirmed in \cite{Kim:2015hya} however, the system of PDEs one had to solve to find a solution was rather difficult to deal with.  Later, utilising some powerful mathematical tools, all local solutions were given in terms of two holomorphic functions (or equivalently a pair of flat space Laplace equations in two dimensions)  in \cite{DHoker:2016ujz}  with some specific physical solutions given in \cite{DHoker:2016ysh,DHoker:2017mds,DHoker:2017zwj} (see also \cite{Corbino:2017tfl}).

Here, which builds on section 5.2 of \cite{Macpherson:2016xwk}, we will be able to give an alternative formulation, where solutions are in one-to-one correspondence with the solutions of a single Laplace equation in two variables. Our starting point is section \ref{sec:mink5system}: First we perform the change of coordinate
\beq
x_1= \tilde x_1+\frac{1}{x_2}e^{2C}B_0
\eeq
which has the effect of re-defining just one element of the vielbein as
\beq
v_1=  e^{3A-\Phi}(d\tilde x_1+ \frac{1}{x_2}e^{2C} H_1),
\eeq
so that the internal metric (from the Mink$_5$ perspective) is of the form
\beq \label{eq:Minint}
d s^2(M_5) =  e^{-6A + 2 \Phi} \bigg( d x_2^2+ x_2^2 d s^2 (S^2)  \bigg) + e^{-2A} d x_3^2+ e^{6A - 2 \Phi} \bigg(d\tilde x_1+ \frac{1}{x_2}e^{2C} H_1\bigg)^2 .
\eeq
Now, similarly to \cite{Macpherson:2016xwk}, we redefine
\beq
\tilde x_1=\frac{1}{2} e^{-3\rho}f(r,y),~~~x_2=\frac{8}{3}e^{3\rho}y,~~~ x_3=4 e^{\rho-\frac{1}{3}\Delta(r,y)} ,~~~ e^{2A}= 2\sqrt{\frac{2}{3}}e^{2\rho+ \frac{1}{2}\Phi+2\lambda}
\eeq
where $e^{\rho}$ will be the AdS radial coordinate, $\Delta$ and $\lambda$ are auxiliary functions, and the  specific powers in $x_i$ are fixed such that they cancel those coming from the $e^{A}$ factors in \eqref{eq:Minint}.
If we demand that the metric, dilaton and NS3-form respect the isometry of $AdS_6$, this leads to 
\begin{align}\label{eq:ads6def1}
e^{\Phi}&= 6 \frac{e^{-2/3 \Delta+4 \lambda}r^2+ f^2 }{e^{8 \lambda}- y^2},~~~ e^{-8\lambda}= \frac{\partial_{y}\Delta}{y(1+ y \partial_{y}\Delta)},\nn\\[2mm]
h_1&=- \frac{e^{-8\lambda-\frac{2}{3}\Delta}y}{9 f}\big(-3r +e^{8\lambda+2/3 \Delta}f \partial_{r}f +r^2 \partial_r \Delta\big),~~~
h_2=-\frac{4}{9}y(\partial_y f- f \partial_{y}\Delta),
\end{align}
where the NS 3-form takes the most general form the isometries allow, namely
\beq\label{eq:ads6def2}
H =( h_1(r,y) dr+ h_2(r,y) dy)\wedge \text{Vol}(S^2).
\eeq
Up to this point we have just imposed conditions leading to AdS$_6$, we need to also impose the supergravity conditions \eqref{eq:Susy2_5d} which lead to
\beq\label{eq:ads6def3}
f=e^{1/3 \Delta}\frac{r\partial_r \Delta-3}{1+ y \partial_y \Delta} , ~~~~ \partial^2_{r}e^{1/3 \Delta}= \frac{1}{3} \partial_{y}^2 e^{-\Delta}.
\eeq
The definitions in \eqref{eq:ads6def1}-\eqref{eq:ads6def3} are sufficient to have an AdS$_6$ solution when the single PDE is solved -the Bianchi identities of the fluxes follow automatically in this case. The metric and fluxes of the solution can be written as
\begin{align}\label{eq:AdS6system1}
 ds^2&=2\sqrt{\frac{2}{3}}e^{ \frac{1}{2}\Phi+2\lambda}\bigg[ds^2(AdS_6)+ \frac{y \partial_y \Delta}{9(1+ y \partial_y \Delta)}ds^2(S^2)+ \frac{\partial_y\Delta}{9 y}\big(dy^2+ e^{-4/3\Delta}dr^2\big)\bigg],\nn\\[2mm]
dC_2&=\frac{2}{81}\bigg(d\left(e^{\frac{1}{3}\Delta-8\lambda}\frac{y^2 \partial_r\Delta}{\partial_y\Delta}\right)- \partial_{y}(e^{-\Delta})dr- 3\partial_{r}(e^{\frac{1}{3}\Delta})dy\bigg)\wedge\text{Vol}(S^2) ,\nn\\[2mm]
H_3&=\frac{4}{3}\bigg(r\big(\partial_{r}(e^{\frac{1}{3}\Delta})dy+\frac{1}{3} \partial_{y}(e^{-\Delta})dr\big)- e^{\frac{1}{3}\Delta}dy -\frac{1}{3} d(y f)\bigg)\wedge\text{Vol}(S^2),\nn\\[2mm]
C_0&=-\frac{1}{18r}\left(1+ 18 e^{\frac{1}{3}\Delta+4\lambda-\Phi}f\right),~~~F_1=dC_0,~~~ F_4=dC_3-C_0 H_3.
\end{align}
In \cite{Macpherson:2016xwk} the T-dual of the unique AdS$_6$ solution in IIA was found as a solution to $\partial^2_{r}e^{1/3 \Delta}= \frac{1}{3} \partial_{y}^2 e^{-\Delta}$, however solving this equation more generally seems difficult. We do however observe a similarity between the system \eqref{eq:AdS6system1} and the classification of AdS$_5\times S^2$ in \cite{Lin:2004nb} (at least in the formers axially symmetric Toda limit).

To proceed we take motivation by the change of coordinates in \cite{Lin:2005nh,Gaiotto:2009gz} (that exploits an idea of \cite{ward}), which maps an axially symmetric Toda equation to the 3d Laplacian in axially symmetric cylindrical coordinates. To simplify the PDE of \eqref{eq:ads6def3} we make an implicit change  of coordinate $(r,y)\to(\eta,\sigma)$ through
\beq
 \sigma =e^{-1/3\Delta},~~~~r= \partial_{\eta}V,~~~~y= \sigma^2 \partial_{\sigma}V.
\eeq
Demanding that the metric is diagonal in $\eta,\sigma$ requires that $V$ satisfies a 4D Laplace equation in  spherically symmetric cylindrical polar coordinates
\beq\label{eq: Vpde}
\frac{1}{\sigma^2}\partial_{\sigma}(\sigma^2 \partial_{\sigma} V)+ \partial_{\eta}^2 V=0,
\eeq
which one can check automatically solves the PDE of \eqref{eq:ads6def3}. We note as a possible point of interest that this is exactly the equation the $SO(3)$ preserving D5-brane of section \ref{sec:D5} obeys.

Given our implicit coordinate change and the definitions \eqref{eq:ads6def1}-\eqref{eq:ads6def3}, it is possible to express every supersymmetric AdS$_6$ solution in IIB in terms of $V$ only as
\begin{align}
 ds^2&=\frac{2\sqrt{2}}{3^{3/4}}e^{ \frac{1}{2}\Phi}\sigma \left(\frac{\Lambda\partial_{\sigma}V}{\partial_{\eta}^2 V}\right)^{1/4}\bigg[ds^2(AdS_6)+ \frac{\partial_{\sigma}V\partial_{\eta}^2V}{3\Lambda}ds^2(S^2)+ \frac{\partial_{\eta}^2V}{3\sigma\partial_{\sigma}V}\big(d\sigma^2+ d\eta^2\big)\bigg]\nn,\\[2mm]
e^{\Phi}&=  \frac{6 \sqrt{3}}{\sqrt{\frac{\Lambda\partial_{\sigma}V}{\partial_{\eta}^2 V}}\partial_{\eta}^2 V}\bigg(3 \big(\Lambda \sigma^2+ (\partial_{\eta}V)^2+2 \sigma\partial_{\eta}V\partial_{\sigma}\partial_{\eta}V\big)\partial_{\sigma}V+ \sigma\big((\partial_{\eta}V)^2-9 (\partial_{\sigma}V)^2\big)\partial_{\eta}^2V \bigg),\nn\\[2mm]
B&= \frac{4}{3}\bigg(\frac{\sigma \partial_{\sigma}V\big(\partial_{\eta}V\partial_{\sigma}\partial_{\eta}V+ \sigma\big[(\partial_{\sigma}\partial_{\eta}V)^2+ (\partial_{\eta}^2V)^2\big]\big)}{\Lambda}- V- \sigma\partial_{\sigma} V\bigg)\text{Vol}(S^2),\nn\\[2mm]
C_0 &=-\frac{1}{18}\frac{3 \partial_{\sigma}V(\partial_{\eta} V+\sigma\partial_{\sigma}\partial_{\eta}V)+\sigma \partial_{\eta}V\partial_{\eta}^2V)}{\sigma (\partial_{\eta}V)^2 \partial_{\eta}^2V+ 3 \partial_{\sigma}V\big((\partial_{\eta}V+\sigma \partial_{\sigma}\partial_{\eta}V)^2+(\sigma \partial_{\eta}^2V)^2\big)},\nn\\[2mm]
C_2&=\frac{2}{27}\bigg(\eta- \frac{\sigma \partial_{\sigma}V\partial_{\sigma}\partial_{\eta}V}{\Lambda}\bigg)\text{Vol}(S^2),~~~~\Lambda=\sigma (\partial_{\eta}\partial_{\sigma}V)^2+ (\partial_{\sigma}V-\sigma \partial_{\sigma}^2V) \partial_{\eta}^2V,
\end{align}
where we have expressed the fluxes in terms of their potentials such that $H= dB,~F_1=dC_0,~ F_3= dC_2- C_0 H$.

Clearly this classification of $AdS_6$ solutions looks quite different to that of \cite{DHoker:2016ujz}, with solutions here governed by \eqref{eq: Vpde} and there by two holomorphic functions. But both classifications should be equivalent, which means there should be a map between the spherically symmetric cylindrical Laplace equation  and holomorphic functions. It would be interesting to find this map and to verify whether such a map exists for all effectively 2 dimensional cylindrical Laplace systems. Another example of interest, which is described a very similar PDE (i.e. a 3d cylindrical Laplace equation with axial symmetry), is the $\mathcal{N}=2$ AdS$_5$ solutions in M-theory studied by Gaiotto and Maldecena in \cite{Gaiotto:2009gz}. In any case in appendix \ref{app: ads6}, we propose a further polar change of coordinates, and a factorization ansatz that generates infinite local solutions with a nontrivial physical region, we also write the metric, dilaton and fluxes for one specific example and discuss the regions where the solution is physical. 

As a final comment we remind that the subclass of the classification in section \ref{sec:mink5system} is more general than just AdS, as it accommodates Minkowski$_5$ solutions representing intersecting $(p,q)$-webs of 5-branes with an AdS$_6$ near-horizon (this is similar to \cite{imamura-D8} for NS5-D6-D8-brane intersections in IIA). It would be interesting to find the complete brane solutions for these systems of $(p,q)$-webs of 5-branes as a further subclass of \ref{sec:mink5system}.

\subsubsection{AdS$_5$}\label{sec:ads5}
Let us see how to recover $AdS_5$ solutions in IIB. We will realise them starting within section \ref{sec:conformalCalab}, which is the only case consistent with $AdS_5$ - specifically, the other cases have warped $U(1)$ factors that cannot lead to a compact internal space.  We will see that imposing $AdS_5$ on our background gives an automatic enhancement of the R-symmetry  to $SU(2) \times U(1)$, consistent with the $\mathcal{N} = 2$ superconformal algebra in four dimensions.

Since we seek an $AdS_5$ solution the parenthesized part of the metric in \eqref{eq: conformalcalmet} should be a cone over some compact space. To achieve this we parametrise the local coordinates as
\begin{equation}
x_1 = r \lambda(\alpha) \cos \beta, \quad x_2 = r  \lambda(\alpha) \sin \beta , \quad x_3 = r  \mu(\alpha) \sin \psi, \quad x_4 = r \mu(\alpha) \cos \psi,
\end{equation}
where $\lambda$ and $\mu$ are arbitrary functions which can be viewed as the radial coordinates of $\R^4$ and $\R^2$ in \eqref{eq: conformalcalmet} respectively, and $r$ is the radial coordinate of our putative $AdS$ factor. 

For an $AdS_5$ the dilaton and NS 3-form  $H$ should have no functional dependence on, or legs in,  $r$. As a consequence the supersymmetry conditions \eqref{NSNS_kperp_z=0} impose
\beq
\label{AdS5_fcond}
H=0,~~~ f=f(\psi).
\eeq
However, we should also impose that the metric contains no $dr$ cross-terms,  which implies
\begin{equation}
\label{AdS5_coordsconstraint}
\lambda(\alpha)^2 + f \mu(\alpha)^2 = c^2
\end{equation} 
where c is a constant. The equation above tells us that $f$ can just be a function of $\alpha$, which is consistent with \eqref{AdS5_fcond} just if $f$ is constant. Without loss of generality one can fix $f = c^2$ and therefore we are free to solve \eqref{AdS5_coordsconstraint} by
\begin{equation}
\mu(\alpha) = \cos \alpha \, , \qquad \lambda (\alpha) = c \sin(\alpha) \, .
\end{equation}
Making this substitution the complex 4D vielbein becomes
\begin{align}
v&=e^{-A-i\beta}\frac{c r}{\sqrt{f}}\bigg( \cos\alpha d\alpha + \sin\alpha\left( \frac{dr}{r} - i d\beta\right)\bigg) ,\nn\\[2mm]
w&=e^{-A - i\psi}r \bigg(i \sin\alpha d\alpha-\cos\alpha\left(i \frac{dr}{r}+ d\psi\right)\bigg).
\end{align}
Comparing this to \eqref{eq: 6dcomplexvielbein} and \eqref{eq: 6dbispinors} we see that the holomorphic 3-form of these solutions depends on $e^{i\psi}$ so that if $\partial_{\psi}$ becomes an isometry, it must be parametrising an R-symmetry. 

$H=0$ means locally $g=0$ and this condition, together with $f = c^2$, turns off almost all the fluxes in \eqref{fluxesConformalCalab}, leaving only $F_5$.

Let us see how the warp factor behaves. It has to satisfy two constraints: one is the Bianchi identity
\eqref{eq:CalabBianchi2} , which now reads
\begin{equation}
\label{AdS_5bianchi}
\Box_{M_6} e^{-4A} = 0 ,
\end{equation}
the other is the $AdS_5$ factorization property, in this case $e^{2A} = L(\alpha,\beta,\psi)^{-2} r^2$. However since the warp factor appears in the metric as $e^{2A}d s^2_{Mink_4}$ and $e^{-2A}d r^2$, we must impose that $L$ is constant and so $\partial_{\psi}$ becomes an isometry. We end up with
\begin{align}
d s_{10}^2 &= \frac{r^2}{L^2}ds^2(\text{Mink}_4)+ \frac{L^2dr^2}{r^2} + L^2\bigg(d \alpha^2 + \cos^2 \alpha d \psi^2 + \sin^2 \alpha (d \beta^2 + \sin^2 \beta d s^2(S^2))\bigg), \nn\\[2mm]
F_5&  = (1 + \star_{10})\text{Vol}_{4} \wedge \frac{d r^4}{L^4} , \quad F_1 = F_3 = B = 0, \quad e^{\phi} = c^{-2} .
\end{align}
So the only AdS$_5\times S^2$ solutions in IIB are locally $AdS_5\times S^5$ where here $S^5$ is parametrised as a foliation of $S^1\times S^3$. That this is all we find is consistent with the result of \cite{Colgain:2011hb}. We can also restrict the period of $\psi$ to a fraction of $2\pi$, \cite{afm}, so that we have an orbifold of $S^5$, and the orbifold singularity is at $\alpha=\pi /2$. The dilaton in this case is fixed by $F_5$ flux quantization  to a nontrivial value depending on the period of $\psi$. Moreover, for these solutions $g_S\geq1$, so they can be seen as non-perturbative F-theory backgrounds \cite{afm}, in fact these are near-horizons of D3-branes in 7-brane singularities, where the 7-branes wrap $S^3 \subset S^5$.
\subsection{Linear dilaton }\label{sec:linearD}
We would now like to see if we can find any solutions with asymptotically linear dilaton. Solutions with linear dilatons are of interest in the context of the holography of  ``Little string theories'' \cite{Berkooz:1997cq} - the field theories on the world volume of a stack of NS5-branes in the decoupling limit. Solutions that interpolate between this behaviour and fractional brane behaviour as one varies the holographic coordinate have been a fruitful avenue for studying holographic duals to confining gauge theories in 4 and 3 dimensions \cite{Maldacena:2000yy,Maldacena:2001pb}. This section should be viewed as preliminary work towards a similar construction in 5d.

Our first task is to  find a simple linear dilation solution within our classification -
we start this with the class of solution in \ref{sec:mink5anstz1}. We demand that only NS3 form flux is turned on and a quick glance at \eqref{eq:mink5anstz1flux}  tells  us we should set $A=\partial_{y_3}E=0$ to do this - this restricts our considerations to case 1, specifically the part that is T-dual to \cite{imamura-D8}. We need to solve only \eqref{eq:masslesspde} which after redefining $(x_1=r \cos\alpha,~x_2=r \sin\alpha)$ becomes
\beq
2\cot\alpha \partial_{\alpha}F + \partial_{\alpha}^2F +\frac{1}{r}\partial_{r}(r^3 \partial_r F)=0.
\eeq
A simple solution to  this PDE is $F=-L\alpha \csc\alpha r^{-1}$ which one can check actually leads to the NS5-brane
\beq
ds^2= ds^2(\text{Mink}_6)+ \frac{L}{r^2}\bigg(dr^2+ r^2ds^2(S^3))\bigg),~~~H= 2L \text{Vol}(S^3),~~~e^{\Phi}=\frac{L}{r^2} ,
\eeq
where $S^3$ is spanned by $(\alpha,S^2)$, which is  indeed a linear dilaton solution up to redefining $r=e^{\rho}$ (albeit a rather trivial one).

We would now like to generalize: As we found the NS5-brane in the ans\"atze of section \ref{sec:mink5anstz1} it seems sensible to stay here, but we shift our attention to case 3 - which has nontrivial D5 and D7 branes in addition to NS5s. Thus we seek  solutions to \eqref{Ceq_subansatz2} and \eqref{Eeq_subansatz2} that avoid simplifying the fluxes too much.
An easy way to solve  \eqref{Ceq_subansatz2} is by setting
\begin{equation}
C =\left(c_3 + \frac{c_4}{x_2}\right) (\partial_{x_3}e(x_3))^2 + l_1 + \frac{l_2}{x_2} \quad \text{and} \quad E = \left(c_3 + \frac{c_4}{x_2}\right) e(x_3) + g(x_2) ,
\end{equation}
where $g$ and $e$ are arbitrary functions on the given support. To find them we can use \eqref{Eeq_subansatz2}. The most general solution is given by
\begin{equation}
\begin{split}
&e = \frac{k_1}{2} x_3^2 + k_2x_3 + k_3 \\
&g = -\frac{k_1 c_3^2}{6} x_2^2 - c_3 c_4 k_1 x_2- c_4^2 k_1 \log(x_2) + \frac{k_4}{x_2} + k_5 .
\end{split}
\end{equation}
The dilaton is then given by $e^{-2 \Phi} = e^{-4A}/\partial_{y_1} F$, which, using our solutions, gives
\begin{equation}
e^{-2 \Phi} = k_1 y_1 + \frac{l_1 + \frac{l_2}{y_2}}{c_3 + \frac{c_4}{y_2}}\,.
\end{equation}
We can notice that, up to a change of coordinates $y_i \rightarrow e^{y_i}$, we have linear dilaton in the following situations:
\begin{itemize}
	\item $l_1 = l_2 = 0 , \quad \Rightarrow \quad e^{-2 \Phi} = k_1 y_1$
	\item $c_3 = l_2 = 0 , \, c_4 = 1 \quad \Rightarrow \quad e^{-2 \Phi} = k_1 y_1 + l_1 y_2$
	\item $k_1 = c_4 = l_1 = 0, \, c_3 = 1 \quad \Rightarrow \quad e^{2 \Phi} = y_2 / l_2$.
\end{itemize}
The metric reads
\begin{equation}
d s^2 = e^{2A} d s^2 (\text{Mink}_5)+  e^{-2A} H_2 \big( d y_2^2+ y_2^2 d s^2 (S^2)  \big) + e^{-2A} d y_3^2+ e^{2A} H_2 \left(d y_1 + H_3 d y_3\right)^2 
\end{equation}
where
\begin{equation}
\begin{split}
&e^{-4A} = k_1 y_1 \left(c_3 + \frac{c_4}{y_2}\right) + l_1 + \frac{l_2}{y_2} \, , \\
&H_2 = c_3 + \frac{c_4}{y_2} \, , \qquad H_3 = k_1 y_3 + k_2 \, .
\end{split}
\end{equation}
And the fluxes are given by
\begin{equation}
\begin{split}
&B = \left[c_4 \left(\frac{k_1}{2} y_3^2 + k_2y_3 +y_1 \right) + k_1 \left( \frac{c_3^2}{3} y_2^3 + c_4 c_3 y_2^2 + c_4^2 y_2  \right) \right] \text{Vol}(S^2)  \\
&F_3 = \left[ k_1 (c_4 + c_3 y_2)^2 (k_2 + k_1 y_3) d y_2 - (l_2 + c_4 k_1 y_1) d y_3 \right] \wedge \text{Vol}(S^2) \\
&F_1 = - k_1 d y_3 , \qquad F_5 = 0 \, .
\end{split}
\end{equation}
We have provided here a new solutions with intersecting NS5-D5-D7, with an asymptotic linear dilaton. 

\section*{Acknowledgments}
We would like to thank Alessandro Tomasiello for various fruitful disscussion throughout this project. F.A. is supported by the NSF CAREER grant PHY-1756996 and by the NSF grant PHY-1620311. The work of J.G. was supported by the Deutsche Forschungsgemeinschaft (DFG) under the grant LE 838/13 and by the Research Training Group RTG 1463 "Analysis, Geometry and String Theory". N.M. has been variously funded by INFN; the European Research Council under the European Union's Seventh Framework Program (FP/2007-2013); ERC Grant Agreement n. 307286 (XD-STRING); and the Italian Ministry of Education under the  Prin project "Non-Perturbative Aspects of Gauge Theories and Strings" (2015MP2CX4).
The work of M.Z. was in part supported by the  the Research Training Group RTG 1463 "Analysis, Geometry and String Theory" of the Deutsche Forschungsgemeinschaft (DFG).

\appendix

\section{$AdS_6$ solutions in polar coordinates} \label{app: ads6}

In this section we look at some examples that follow from solving cylindrical Laplace equations \eqref{eq: Vpde}, which is the only PDE that regulates the AdS$_6$ solutions of IIB. To do so we perform a further change of coordinates,
\begin{equation}
\sigma = \rho \cos (\omega), \qquad \eta= \rho \sin (\omega),
\end{equation}
mapping the plane $\{ \sigma, \eta\}$ to a disc. Equation \eqref{eq: Vpde} transforms in the following way
\begin{equation}
\frac{1}{\rho^2}\left(2 \cot (\omega) \partial_{\omega} V + \partial^2_{\omega} V+ 3 \rho \partial_{\rho} V\right)+ \partial_{\rho}^2 V =0.
\end{equation}
In order to solve this equation we implement a factorization ansatz,
\begin{equation}
V(\rho, \omega) = f(\rho) u(\omega).
\end{equation} 
This also factorizes \eqref{eq: Vpde} in two decoupled ODEs:
\begin{subequations} \label{eq: ODEro}
	\begin{align}
	& \rho (3f'(\rho)+ \rho f''(\rho))+ s f(\rho)=0\\
	& 2 \cot (\omega) u'(\omega)+ u''(\omega) - s u(\omega)=0,
	\end{align}
\end{subequations}
for a constant $s \in \mathbb{R}$. This generates an infinite class of solutions depending on 5 parameters $\{s, c_1, c_2, \widetilde c_1, \widetilde c_2\}$, which are not always independent. The general solutions of the system \eqref{eq: ODEro} highly depend on the value of $s$:
\begin{itemize}
	\item for $s>1$ they read
	\begin{align}
	&f(\rho)=\frac{c_1 \sin (\sqrt{s-1}\log(\rho))+c_2 \cos (\sqrt{s-1}\log(\rho))}{\rho},\\
	&u(\omega)=\frac{\widetilde c_1 e^{-\sqrt{s-1} \omega} +\widetilde c_2 e^{\sqrt{s-1} \omega}}{\sin(\omega)};
	\end{align}
	\item for $s=1$ they read
	\begin{align}
	&f(\rho)=\frac{c_1 +c_2 \log(\rho)}{\rho},\\
	&u(\omega)=\frac{\widetilde c_1 +\widetilde c_2  \omega}{\sin(\omega)};
	\end{align}
	\item for $s<1$ and $s\neq 0$ they read
	\begin{align}
	&f(\rho)=\rho^{-1 - \sqrt{1-s}}\left(c_1+c_2 \rho^{2 \sqrt{1-s}}\right),\\
	&u(\omega)=\frac{2\widetilde c_1 \cos(\sqrt{1-s} \, \omega)}{\sin(\omega)}
	\end{align}
	with $\widetilde c_2 = i \widetilde c_1$, and alternatively 
	\begin{equation}
	u(\omega)=\frac{2\widetilde c_2 \sin(\sqrt{1-s} \, \omega)}{\sin(\omega)}
	\end{equation} 
	with $ \widetilde c_1 = i \widetilde c_2$;
	\item at last for $s=0$ we have
	\begin{align}
	&f(\rho)=\left(-c_1 \rho^{-2}+c_2\right),\\
	&u(\omega)=\widetilde c_2;
	\end{align}
	with $\widetilde c_1 = \frac{i}{2} \widetilde c_2$, or alternatively 
	\begin{equation}
	u(\omega)=2 \widetilde c_1 \cot(\omega)
	\end{equation}
	with $\widetilde c_2 = 2 \widetilde c_1$.
\end{itemize}
As an example, we will explicitly calculate fluxes and metric for this last case. We change the name of the constants, $\tilde c_i \mapsto c_i$, so that:
\begin{equation}
V = c_1 + \frac{c_2}{\rho^2}.
\end{equation}
The positivity of all the square roots in the metric imposes the following conditions on the angular coordinates $\omega$:
\begin{equation}
\cos \omega > 0 , \qquad -1 + 2 \cos(2 \omega) > 0 \\
\end{equation}
which implies that we must restrict ourselves to the region $\omega \in \left( - \frac{\pi}{6} , \frac{\pi}{6} \right)$.
In this situation the metric reads
\begin{equation}
ds^2 = k \left( \frac{12 \cos^2 \omega}{2 \cos(2\omega)-1} ds^2(AdS_6) + ds^2(S^2) + \frac{4}{\rho^2}(d \rho^2 + \rho^2 d \omega^2) \right)
\end{equation}
where
\begin{equation}
k = \frac{2 |c_1|}{3} \frac{\cos \omega(2 \cos(2\omega)-1)}{\rho^{3/2}} \sqrt{\frac{c_1}{|c_1|} \frac{1- 12 \rho \cos^3 \omega + 8 \cos(2 \omega)}{\cos(3 \omega)}}.
\end{equation}

We can notice that we have another constraint which comes from the radial coordinate, indeed we must have $1- 12 \rho \cos^3 \omega + 8 \cos(2 \omega) \lessgtr 0$ depending on the sing of $c_1$: if $c_1 < 0$ then $\rho$ is bounded from below while if $c_1 > 0$ then $\rho$ is bounded from above.

The fluxes read:
\begin{equation}
C_0 = \frac{2 \rho^3 \sin\omega}{9c_1(2 \cos(2 \omega)-5)} , \quad C_2 = \frac{4 \rho\sin \omega}{27} \text{Vol}(S^2) , \quad B_2 = \frac{2 c_1 (4 \cos(2 \omega) - 3)}{3 \rho^2} \text{Vol}(S^2) 
\end{equation}
while the dilaton is:
\begin{equation}
e^{\Phi} = 6 \sqrt{3} c_1 \cos \omega \frac{1- 12 \rho \cos^3 \omega + 8 \cos(2 \omega)}{\rho^2 \sqrt{2 \cos(2\omega)-1}} .
\end{equation}

\end{document}